\documentclass[preprint,preprintnumbers,amsmath,amssymb, superscriptaddress,letterpaper, prb]{revtex4}
\usepackage{graphicx}
\usepackage{dcolumn}
\usepackage{subfigure}
\usepackage{bm}

\newcommand{\vk}{V({\bf k})}

\newcommand{\sumj}{\sum_{j>i}^N}
\newcommand{\br}{{\bf r}}
\newcommand{\bk}{{\bf k}}

\newcommand{\rhok}{\rho({\bf k})}
\newcommand{\rhonk}{\rho({\bf -k})}

\begin{document}

\preprint{}

\title{Inherent Structures for Soft Long-Range Interactions in Two-Dimensional Many-Particle Systems}
\author{Robert D. Batten}
\affiliation{Department of Chemical Engineering, Princeton University,
Princeton, NJ 08544, USA}

\author{Frank H. Stillinger} 
\affiliation{Department of Chemistry, Princeton University, Princeton, NJ, 08544
USA}

\author{Salvatore Torquato} 
\affiliation{Department of Chemistry, Princeton University, Princeton, NJ, 08544
USA} 
\affiliation{Princeton Center for Theoretical Science, Princeton University,
Princeton, NJ 08544, USA}
\affiliation{Program in Applied and Computational Mathematics, Princeton
University, Princeton, NJ 08544, USA} 
\affiliation{Princeton Institute for the Science and Technology of Materials,
Princeton University, Princeton, NJ 08544, USA}

\altaffiliation{Corresponding author}
\email{torquato@electron.princeton.edu} \date{\today}

\date{\today}

\begin{abstract}
We generate inherent structures, local potential-energy minima, of the
``$k$-space overlap potential'' in two-dimensional many-particle systems using a
cooling and quenching simulation technique. The ground states associated with
the $k$-space overlap potential are stealthy ({\it i.e.,} completely suppress
single scattering of radiation for a range of wavelengths) and hyperuniform
({\it i.e.,} infinite wavelength density fluctuations vanish). However, we show
via quantitative metrics that the inherent structures exhibit a range of
stealthiness and hyperuniformity depending on the fraction of degrees of freedom
that are constrained. Inherent structures in two dimensions typically contain
five-particle rings, wavy grain boundaries, and vacancy-interstitial defects.
The structural and thermodynamic properties of inherent structures are
relatively insensitive to the temperature from which they are sampled,
signifying that the energy landscape is relatively flat and devoid of deep
wells. Using the nudged-elastic-band algorithm, we construct paths from
ground-state configurations to inherent structures and identify the transition
points between them. In addition, we use point patterns generated from a random
sequential addition (RSA) of hard disks, which are nearly stealthy, and examine
the particle rearrangements necessary to make the configurations absolutely
stealthy. We introduce a configurational proximity metric to show that only
small local, but collective, particle rearrangements are needed to drive initial RSA
configurations to stealthy disordered ground states. These results lead to a
more complete understanding of the unusual behaviors exhibited by the family of
``collective-coordinate'' potentials to which the $k$-space overlap potential
belongs.
\end{abstract}

\maketitle

\section{Introduction}
\label{sec:intro}
Recently, we have been interested in a family of soft, long-range pair
interactions that give rise to novel physical behaviors of many-particle systems
including classical disordered ground states for a range of densities, negative
thermal expansion, and vanishing normal-mode frequencies.\cite{uche2006ccc,
uche2006ccc, torquato2008ndr, batten2008cdg, batten2009novel,
batten2009interactions, torquato2009iot} This family of soft pair interactions
includes those pair potentials $v(r)$ with Fourier transforms $V(k)$ that are
positive, bounded, and vanish to zero at some finite wavenumber $K$ and beyond. 
The potential energy $\Phi$ of a system of $N$ particles in a fundamental cell
with volume $\Omega$ under periodic boundary conditions can be written as
\begin{eqnarray}
\label{eq:phi}
 \phi  = \Phi/N &=& \frac{1}{N}\sumj v(\br_{ij}), \\ 
                &=&  \frac{1}{2\Omega} \sum_{\bk} \vk [\rhok\rhonk/N - 1], 
\end{eqnarray} 
where $r_{ij}\equiv |\br_i-\br_j|$ is the distance between particles $i$ and
$j$, $\rhok$ are the Fourier coefficients of the density field, and $\bk$ are
the wave vectors appropriate for the system size and shape. These soft pair
interactions, called ``collective-coordinate'' potentials, are bounded and
possess long-range oscillations in real space.\cite{torquato2008ndr}

Because of the finite cutoff in $k$-space, an analytic lower bound for the
potential energy per particle $\phi$ is easily obtainable, and numerical methods
allow for the construction and investigation of ground states with extremely
high precision. A series of studies has examined the structural characteristics
of ground states in one,\cite{fan1991ccd} two,\cite{uche2004ccd} and three
dimensions.\cite{uche2006ccc} Three structural ground-state regimes were found
to characterize these two-dimensional systems. Increasing the fraction of
degrees of freedom that are constrained $\chi$, equivalently decreasing density,
spans regimes that are disordered, wavy-crystalline, and crystalline. Recently,
an analytical connection between the fraction of constrained degrees of freedom
within the system and the disorder-order phase transition for a class of target
structure factors has been provided\cite{zachary2011inpress} by examining the
realizability of the constrained contribution to the pair correlation function.

\begin{figure}
\subfigure[ $\text{ }$Conventional]
{\label{fig:normglass}\includegraphics[width=0.4\textwidth, clip=true]{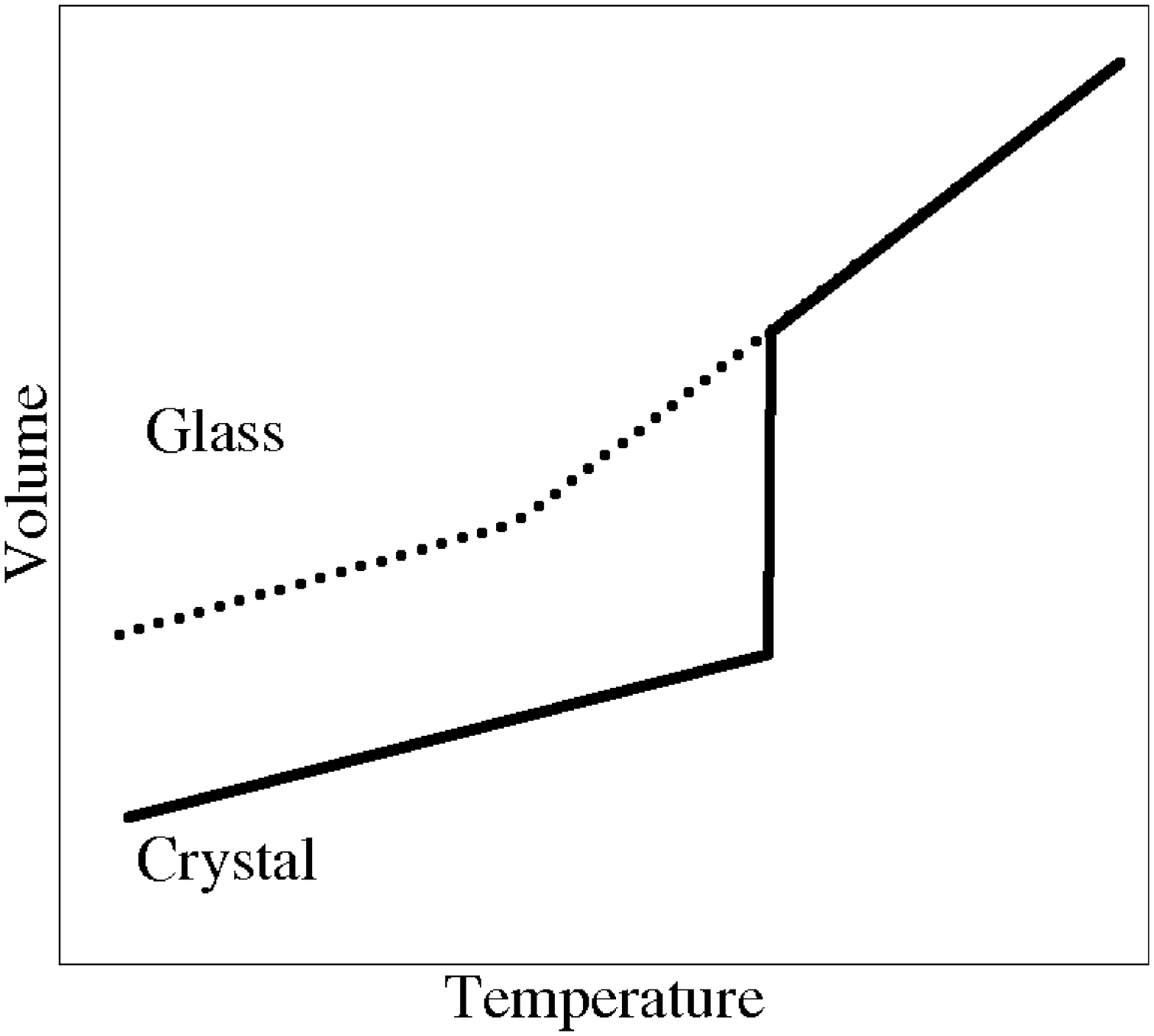}}
\subfigure[ $\text{ }$CollectiveCoordinate]
{\label{fig:ccglass}\includegraphics[width=0.4\textwidth,clip=true]{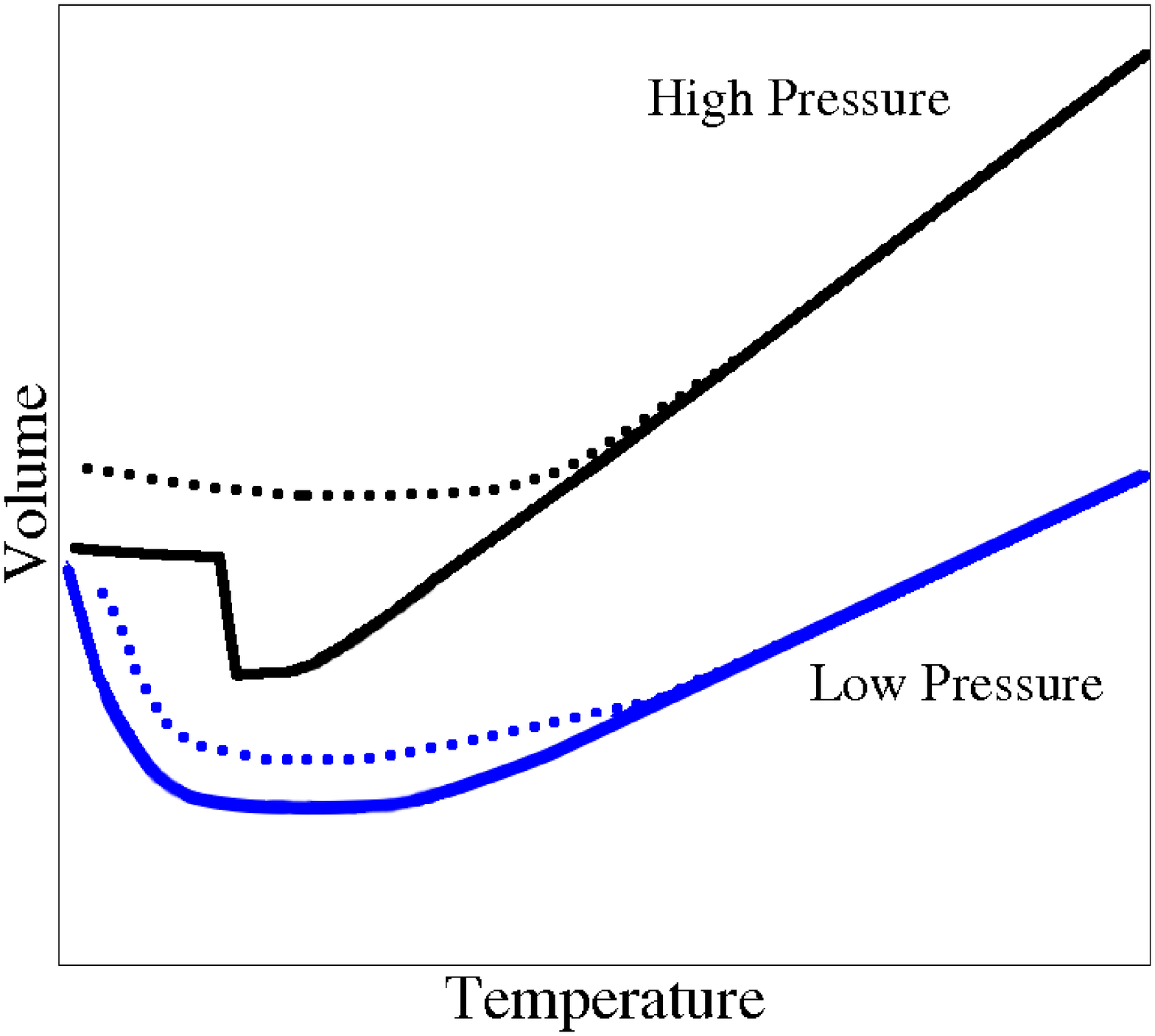}}
\caption{(Color online) Schematics of the equilibrium equation of state (solid
lines) and the nonequilibrium path that results in glassy behavior (dotted
lines) along an isobar. With familiar potentials ({\it e.g.} Lennard-Jones)
(left), the equilibrium curve shows a first-order phase transition while the
glassy curve undergoes supercooling and a glass transition.  With the soft, long
range $k$-space overlap potential (right), high-pressure systems result in a
first-order transition that is not present for low-pressure systems. Both
collective-coordinate systems undergo negative thermal expansion.}
\label{fig:glasses}
\end{figure}

More recently, these potentials were used to construct ``stealthy'' and
``hyperuniform'' materials in two dimensions.\cite{batten2008cdg} ``Stealthy''
materials refer to point patterns that completely suppress single scattering of
radiation for certain wavelength ranges.  While nonstealthy materials may allow
for some small amount of scattering, stealthy materials are absolutely
transparent at those wavelengths.  ``Hyperuniform'' refers to point patterns in
which infinite wavelength density fluctuations vanish.\cite{torquato2003ldf} 
(These terms are defined more precisely in Sec.\ \ref{sec:background}.)  The
hyperuniformity notion enables the rank ordering of crystals, quasicrystals and
special disorderedmany-particle systems.\cite{torquato2003ldf,
zachary2011statmech, zachary2011prl} Interestingly, disordered hyperuniform
many-particle ground states, and therefore also point distributions, with
substantial clustering can be constructed.\cite{zachary2011inpress}

Constructed ground states are now the basis for novel disordered materials with
tunable, photonic band gaps.\cite{florescu2009designer,man2010experimental}
Using a model potential from the family of collective-coordinate potentials in
two dimensions, the ``$k$-space overlap potential,'' described in Sec.\
\ref{sec:background}, we observed negative thermal expansion and vanishing
normal-mode frequencies for ground-state configurations. We attributed these
phenomena to the nature of the underlying energy landscape, which we described
as having ground-state ``valleys'' that weave between the higher-energy portions
of the landscape.\cite{batten2009novel, batten2009interactions}

Despite the discovery of the existence of such unique properties, the
fundamental mechanisms allowing for disordered ground states and stealthy point
patterns still need to be fully elucidated. Here, we analyze the energy
landscape further by examining the associated inherent structures, or local
potential-energy minima, for the aforementioned $k$-space overlap potential in
two dimensions to better understand these properties.

We have two primary motivations for this work. First, through the course of our
previous research, we have observed several paradoxical phenomena related to
local potential-energy minima and glassy behavior.  While an ``inherent
structure'' is in general any local potential-energy minimum, the term ``glass'' refers
to an amorphous solid that is kinetically trapped in a potential-energy well. 
In Ref.\ [2], we observed an unusual equation of state when cooling a system
compared to a more familiar ({\it e.g.}, Lennard-Jones) glass-forming system as
Fig.\ \ref{fig:glasses} demonstrates this schematically.  In Fig.\
\ref{fig:normglass}, the dark and dotted lines represent the equilibrium
equation of state and nonequilibrium isobaric cooling path for a Lennard-Jones
system. Figure \ref{fig:ccglass} shows the corresponding cooling paths for a
system interacting with the $k$-space overlap potential.  In the more familiar
glass-forming system, there is a well-defined glass transition below the
freezing point that is dependent on the rate at which the system is
cooled.\cite{debenedetti1996metastable}  It is well known that the glass-forming
behavior of a system is a function of the underlying energy
landscape.\cite{goldstein1969viscous} The depths of the potential-energy wells
in the energy landscape dictate the dynamics of structural rearrangements of the
system.\cite{sastry1998signatures, debenedetti2001supercooled} For the $k$-space
overlap potential, we observed that at high pressure the nonequilibrium curve
deviated from the equilibrium curve at a temperature above the melting
temperature, as demonstrated in Fig.\ \ref{fig:ccglass}.  At lower pressures
where there is no well-defined melting temperature, as in the wavy-crystalline
regime, the nonequilibrium curve deviates similarly from the equilibrium curve.
In addition, the $k$-space overlap potential gives rise to negative thermal
expansion, generally for nondimensional $T^*<0.0007$.\cite{batten2009novel,
batten2009interactions}

Other paradoxical behaviors have led us toward a broad analysis of the energy
landscape. We have observed that some ground-state configurations, particularly
crystalline structures, had many zero-energy normal modes. These ground states
are therefore not mechanically rigid. However, while using numerical
minimization algorithms to construct ground-state configurations, we encountered
many mechanically stable local potential-energy minima.  Lastly, there exists a
large density range in which the energy landscape was evidently devoid of all
local minima. An increment in $\chi$ just outside this range introduced local
potential-energy minima to the landscape.

While we have previously observed these
phenomena,\cite{batten2008cdg,batten2009novel, batten2009interactions} the
fundamental mechanisms that underlie them are not fully understood.  The
inherent-structure analysis has proven to be a fruitful method for relating the
energy landscape to low-temperature phenomena.  Many studies have examined
inherent-structure characteristics in glass-forming liquids with strong
repulsive cores such as the Stillinger-Weber
potential,\cite{stillinger1983dynamics, stillinger1984point,
heuer1997properties} water-like pair potentials,\cite{stillinger1983inherent}
binary Lennard-Jones-like
systems,\cite{weber1985local,weber1985interactions,buchner1999potential,
broderix2000energy} and general repulsive
potentials.\cite{oligschleger1999collective,la2003test} 
The inherent structures for Lennard-Jones and steeply repulsive potentials
are in general not hyperuniform or stealthy due to the dominance of grain boundaries
and vacancy defects.

In contrast to the above list of strongly repelling potentials, the $k$-space
overlap potential is a soft interaction. Soft interactions are often useful
models for soft-matter systems such as colloids, polymers, and
microemulsions.\cite{likos2001eis} In addition, these $k$-space overlap
interactions are also qualitatively similar to Friedel oscillations in molten
metals.\cite{ashcroft1976ssp}  The $k$-space overlap potential is localized in
$k$-space and delocalized in real space as a result of the Fourier transform.
Certain duality relations link the ground-state energies of the $k$-space
overlap potential to the ground-state energies of real space
analog.\cite{torquato2008ndr,torquato2011duality} There have been several
investigations of the inherent structures of various soft interactions.  The
Gaussian-core model in two dimensions has polycrystalline inherent structure
with a large correlation length even when sampled from the liquid
state.\cite{stillinger1982hidden} Energy landscape analyses revealed that the
range of the Morse potential affects the relation between temperature and the
potential-energy distribution of sampled inherent
structures.\cite{shah2002potential}  While for the Yukawa potential, inherent
structures varied depending on whether they were obtained from the liquid,
crystal or hexatic phase.\cite{qi2010melting}

Novel, stealthy dielectric materials are currently being
fabricated,\cite{man2010experimental} and nearly stealthy
 ceramic materials are of interest for optical
applications.\cite{mattarelli2007ugc,mattarelli2010transparency} Because of
these recent experimental applications and the unusual physical behaviors
discovered,\cite{batten2008cdg,batten2009interactions, batten2009novel} we are
also motivated to understand the fundamental differences between point patterns
that absolutely suppress scattering for certain wavelengths and those that
nearly suppress scattering. Our ground-state construction procedure
\cite{batten2008cdg} automatically distinguishes stealthy and hyperuniform
configurations from those that do not possess such properties. In particular, we
want to understand whether the particle rearrangements required to transform
nearly hyperuniform and stealthy materials to configurations that are perfectly
hyperuniform and stealthy are global or local in nature.

While a general method for understanding these particle rearrangements would
require a new algorithm to search for collective motions that increase
stealthiness, the collective-coordinate approach provides an excellent framework
from which to address this question.  Here, we provide two methods for
identifying rearrangements in particle systems to achieve stealthy and
hyperuniform materials. For large $\chi$, we use the nudged-elastic-band
algorithm to connect inherent structures, which are nearly hyperuniform and
nearly stealthy, to ground states along a minimum-energy path. For small $\chi$
values, we have identified that there are no inherent structures higher up in
the energy landscape.  Therefore, we study the rearrangements from a saturated
random sequential addition (RSA) of hard disks\cite{torquato2006rsa} to ground
states of the $k$-space overlap potential. We use saturated RSA systems as
initial conditions because upon saturation they suppress scattering for small
wavenumbers and are nearly hyperuniform.\cite{torquato2006rsa,
mattarelli2007ugc, mattarelli2010transparency} We also introduce a stealthiness
metric and a configurational proximity metric to quantify characteristics of
these transitions.

In this paper, we use a collective-coordinate potential and a simulation
methodology to find stealthy point patterns and local potential-energy minima
above the ground state. We probe the following fundamental questions:
\begin{itemize}
\item How are inherent structures and their thermodynamic properties in
collective coordinate systems different from those found in other soft-potential
systems?
\item To what extent are inherent structures stealthy and/or hyperuniform? 
\item How are the features of inherent structures related to the pair potential
function?
\item What collective particle rearrangements are necessary to construct a path
from an inherent structure to a ground state? 
\item What global and/or local particle rearrangements are necessary to convert
a nearly stealthy and nearly hyperuniform system in to one that is absolutely
stealthy and hyperuniform?
\end{itemize}
The remainder of this paper is organized as follows. In Sec.\
\ref{sec:background}, we define the $k$-space overlap potential, introduce
several definitions, and briefly review the ground-state structural regimes. In
Sec.\ \ref{sec:methods}, we detail the methods we use to obtain and characterize
inherent structures. In addition, we discuss the methods we use to find
transition states between inherent structures and ground states.  In Sec.\
\ref{sec:cooling}, we examine the thermodynamic properties of inherent
structures as a function of system size and cooling rate. The inherent
structures are characterized in Sec.\ \ref{sec:structure}. Paths connecting
inherent structures to ground states are provided in Sec.\ \ref{sec:mapping}
while the rearrangements from RSA systems to ground states are explored in Sec.\
\ref{sec:rsa}. Concluding remarks are provided in Sec.\ \ref{sec:disc}.

\section{Collective Coordinates and the Overlap Potential}
\label{sec:background}

While a fully detailed summary of collective coordinates is provided in Ref.\
[2], we provide a brief summary of the relevant mathematical relations and
definitions here. Recall that the potential energy of the system of interacting
particles in a periodic simulation box is defined in Eq.\ (\ref{eq:phi}). The
collective density variables $\rhok$ are given by
\begin{equation} 
\label{eq:rhok} \rho({\bf k }) = \sum_{j=1}^{N} \exp(i{\bf k \cdot r}_j),
\end{equation}
where ${\bf r}_j$ is the position of the $j^{th}$ particle. The wave vectors for
a periodic box correspond to linear combinations of the reciprocal lattice
vectors associated with the periodic box. For a rectangular box of dimensions
$L_x$ and $L_y$, those wave vectors are $\bk = [ 2\pi n_x/L_x, 2\pi n_y/L_y]$,
where $n_x$ and $n_y$ are integers.  The structure factor $S(\bk)$, proportional
to the intensity of scattering of radiation, is related to the collective
density variables via
\begin{equation}
\label{eq:sk}
S({\bf k}) = \frac{\left<|\rho({\bf k})|^2\right>}{N}, 
\end{equation}
where $<\cdots>$ is the appropriate ensemble average. When angularly averaged,
the structure factor $S(k)$ is only dependent on the wavenumber $k \equiv
|\bk|$. Since the collective-coordinate class of potentials are bounded
positive for $k \le K$ and zero  for $k>  K$, any configuration
for which  $S(k)$ is constrained to be zero for all $0< |\bk| < K$ is a ground
state.

For such potentials, it is useful to refer to the dimensionless parameter 
\begin{equation}
\chi = \frac{M(K)}{dN} \leq 1
\end{equation}
as the fraction of degrees of freedom that are constrained, where $M(K)$ is the
number of independent wave vectors $0<|\bk|<K$ and $dN$ is the total degrees of
freedom.\cite{uche2004ccd}  In two dimensions $\chi$ is limited to the range
$\chi < 0.91$.  This limit was discussed in Ref.\ [2] and arises from the
inability to suppress Bragg scattering of the triangular
lattice.\cite{suto2005cgs}  Here, we fix $K=1$, and therefore $\chi$ is
inversely related to the number density $\rho = N/\Omega$. Some relations
between $\chi$ and $\rho$ are given in Ref.\ [2].  Three structural regimes
exist for ground-state configurations in two dimensions.  For $\chi<0.577$,
ground states are disordered. For $0.577 \leq \chi < 0.780$, they are
wavy crystalline, while for $0.780 \leq \chi \leq 0.91$, crystalline structures
are the most viable ground-states configurations identified via numerical search
techniques.\cite{uche2004ccd, batten2009interactions}  We have observed that
quenching systems with $\chi<0.5$ always resulted in a ground state. The energy
landscape is evidently devoid of local potential-energy minima above the ground
state when $\chi<0.5$ and thus inherent-structure analysis is limited to
those systems with $\chi\geq0.5$.

\begin{figure}
\includegraphics[width=0.8\textwidth, clip=true]{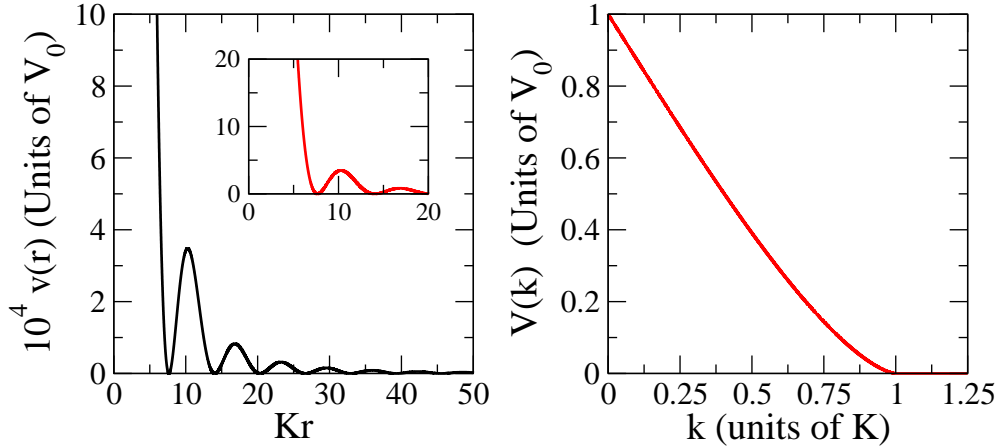}
\caption{The overlap pair potential function $v(r)$ in the infinite-volume limit and its
corresponding Fourier transform $V(k)$, the $k$-space overlap potential, 
as adapted from Ref.\ [2].}
\label{fig:vkvr} 
\end{figure}

We continue to use the $k$-space overlap potential, introduced by Torquato
and Stillinger, \cite{torquato2008ndr}
since it is exactly representable in real and reciprocal space, and
is sufficiently short-ranged in real space 
for computational purposes; see Fig. \ref{fig:vkvr}. The $k$-space overlap potential
is proportional to the intersection area between two disks of diameter $K$ with
centers separated by $k$, and hence for $k \le K$ is
\begin{equation}
\label{eq:overlapvk}
V(k) = \frac{2V_0}{\pi}\left[ \cos^{-1}\left(\frac{k}{K}\right) -
\frac{k}{K}\left(1-\frac{k^2}{K^2}\right)^{1/2}\right],
\end{equation}
and zero for all $k > K$.  In the infinite-volume limit, the associated real space pair potential is
\begin{equation} 
\label{eq:overlapvr} 
v(r) = \frac{V_0}{\pi r^2}\left[J_1\left(\frac{Kr}{2}\right)\right]^2,
\end{equation}
where $J_1$ is the Bessel function. Henceforth, we refer to this $v(r)$ as the
``overlap potential.'' The potential $v(r)$ is bounded at $r=0$ and behaves as
$\cos^2(Kr/2-3\pi/4)/r^3$ for large $r$.

In this paper, we choose to report a scaled potential energy $\varepsilon$,
omitting the structure-independent contribution at $\bk = 0$ and removing the
additive constants, i.e.,
\begin{equation}
\label{eq:scaledphi}
\varepsilon = \frac{1}{2\Omega}\sum_{\bk \neq 0}V(\bk)S(\bk), 
\end{equation}
so that ground state energies are equal to zero regardless of $\chi$ for
$\chi<0.91$. This new scaled energy provides a sense as to how
much a configuration differs energetically from a ground state, and is
related to the actual potential energy, choosing $V_0=1$, via
\begin{eqnarray}
\phi = \Phi/N &=& \frac{1}{\Omega N} \sum_{\bk} \left[V(\bk)C(\bk) \right] \\
&=& \frac{\rho}{2} - \frac{1}{2\Omega}\sum_{\bk}V(\bk) + \varepsilon
\end{eqnarray}
Note that the structure-independent terms sum to be nonnegative because $v(r)$
is a nonnegative function.  We choose the cutoff for ground states to be those
configurations in which $\varepsilon<10^{-10}$.

We also compute a stealthiness metric $\eta$ as the average of the values of the
structure factor for all $|\bk|<K$, defined to be
\begin{equation}
\eta = \sum_{0 < |\bk|<K} S(\bk) / M(K)
\end{equation}
This metric is a measure of the stealthiness because it vanishes to zero for
completely transparent systems but will remain nonzero for nonstealthy systems. 
While there is a numerical limitation of $10^{-16}$ for double precision
computing, there are certain practical limits on stealthiness. Because of the
weights assigned by $V(k)$ for $k$ near $K$, the metric $\eta$ may not
necessarily be suppressed to zero for ground states due to numerical precision
of the quantity $V(k)S(k)$. Therefore, the stealthiness metric is best used as a
relative comparison of stealthiness between configurations and not as a
determination of a configuration as a ground state.

\section{Methods}
\label{sec:methods}

Our method to generate inherent structures follows that of Sastry and
coworkers,\cite{sastry1998signatures} which is based on the initial algorithms
of Stillinger and Weber.\cite{stillinger1983dynamics,stillinger1984point,
weber1985interactions}  Initially, we simulate the atomic motions of the system
interacting via the $k$-space overlap potential in Eq.\ (\ref{eq:phi}) by
integrating the Newtonian equations of motions, assuming unit mass,
\begin{equation}
\frac{d^2\br_i}{dt^2} = -\nabla\phi.
\end{equation}
We use the velocity Verlet algorithm\cite{verlet1967cec} with a time step
$\delta t$ of 0.4, which is chosen so as to accurately conserve total energy in
the constant $NVE$ ensemble.  Throughout the dynamical simulation,
configurations were sampled and subjected to a quenching of the potential
energy.  Here, the sampled configuration is used as the initial condition for a
conjugate gradient minimization.

Upon termination of the conjugate gradient quenching, we subject the system to
an additional quenching via the MINOP
algorithm.\cite{dennis1979tnu,kaufman1999rsq}  We previously reported the
efficiency of the MINOP algorithm compared to the conjugate gradient
method.\cite{uche2006ccc, batten2008cdg} We add this final quench for reasons of
increased efficiency and precision compared to simply using the conjugate
gradient method to high precision.  Many of the inherent structures that we have
found have very slight differences in the scaled potential energy $\varepsilon$
that can be on the order of magnitude $10^{-8}$. Our implementation of the
conjugate gradient method becomes inefficient at these tolerances. While there
is a possibility that the MINOP algorithm can traverse to other capture basins,
we suspect that the system is quenched sufficiently deep into a potential-energy
well after the conjugate gradient step that the MINOP algorithm will not push
the system into another capture basin.

We initialize the dynamical trajectories at a temperature associated with the
liquid state. Here, we report the dimensionless temperature $T^* = k_BT/V_0$,
where $k_B$ is the Boltzmann constant and $T$ is the simulation temperature. We
previously reported that the transitions from ground states to highly disordered
liquids occur in the range of $T^*=0.0003$-$0.001$, depending on
$\chi$.\cite{batten2009interactions} We initialize our simulations at
$T^*\geq0.0018$. Configurations are sampled every $n_s$ time steps until at
least fifty are obtained. Then the velocities of the system are rescaled to a
lower temperature by $dT^*$, typically a value of 0.00015, and the
configurations are sampled again at the reduced temperature.  The rate at which
the system is effectively cooled becomes $\gamma = dT^*/(50n_s\delta t)$.  We
have explored a range of coolings rates from $10^{-6} \leq \gamma \leq 10^{-9}$
with $n_s$ varying from 3 to 750. Surprisingly, as discussed in Sec.\
\ref{sec:cooling}, while the properties of the thermal structures are sensitive
to the cooling rate, the properties of the inherent structures are not.

To construct pathways between ground-state configurations and inherent
structures for $\chi>0.5$, we use the ``nudged-elastic-band algorithm'' used for
finding minimum-energy transition
paths.\cite{henkelman2000improved,henkelman2002methods}  This method requires as
an input two local minima that are nearby in the energy landscape.  We generate
this pair of structures by initializing a known ground-state configuration
(obtained by the search methods in Ref.\ [3]), assigning a small temperature to
the system, and integrating the equations of motion.  The configurations are
quenched by the conjugate gradient/MINOP minimization every five time steps. A
simulation run is terminated when the quenching of a configuration produces a
local minimum in a new capture basin. This new inherent structure is presumed to
be near the original ground state in configuration space.

We then apply the nudged-elastic-band (NEB)
algorithm\cite{henkelman2000improved, henkelman2002methods} to the ground
state/inherent structure pair. The algorithm discretizes the path between two
local minima in a potential-energy landscape into ``beads,'' or image
configurations. These beads, representing configurations, are connected via
harmonic springs of zero natural length. Then a minimization algorithm is
applied to minimize simultaneously the force parallel to the string of beads and
the force perpendicular to the true force. The algorithm bends the string of
beads around the ``hills'' in the energy landscape. The end result is a discrete
path that provides the minimum-energy path from one potential-energy minimum,
through a low-order saddle point, and to the other potential-energy minimum.
While a large number of ``beads'' allows for a more refined minimum-energy
pathway, we find that using fifty beads serves our purposes well by yielding a
smooth transition pathway and a reasonable approximation to the saddle point.

Occasionally, there are other inherent structures in the vicinity of the
straight-line interpolation between the inherent structure and ground state used
as an initial condition. In these cases, the NEB algorithm will not produce a
smooth transition from the ground state to the original inherent structure, and
therefore we refine our initial ground state/inherent structure pair. For
example, when using the triangular lattice as an initial condition for say
$\chi=0.6004$, there are number of zero-energy modes that the system can
traverse during the molecular dynamics trajectories. When applying the NEB
algorithm to a triangular lattice/inherent structure pair, the pathway has a
preference to bend toward the zero-energy path before moving uphill toward the
saddle point.  This caused the energy of the transition-state pathway to appear
rugged since the string of beads was pushed near a different inherent structure.
 In these cases, we searched the string of beads for a new ground-state/inherent
structure pair. We minimized the potential energy of each bead until we found a
new ground state/inherent structure pair closer together than the original pair
and applied the NEB algorithm again.  This method does not guarantee that we
find the closest inherent structure to a ground state nor does it guarantee that
the saddle point is a first-order (single negative eigenvalue) saddle.

For $\chi<0.5$, we examine particle rearrangements from RSA configurations to
ground states.  To generate RSA packings, disks of diameter $D$ are randomly,
sequentially, and irreversibly placed inside a simulation cell so that they do
not overlap with any other particles. The process is saturated when additional
particles cannot be placed without overlapping other particles. In practice,
this occurs after a very large number of attempted particle insertions are
rejected. Here, we terminate the process after 1.5 million attempted particle
placements.  Since $K$ is fixed to be unity, we must select a target number
density $\rho_t$ that corresponds to our target $\chi$. Using a desired number
of particles in a saturated system of 750 particles, we then identify the
appropriate system area. The simulation cell is chosen to be a square box. We
can then assign the particles a diameter $D$ so that upon saturation of the RSA
process, we achieve close to the desired number density. For RSA processes in
two dimensions, the saturation packing fraction $f_s$ is
0.547.\cite{torquato2006rsa} The assigned diameter becomes $D =
2(f_s/\pi\rho_t)^{1/2}$.  In our RSA patters, the number of particles did not
match the target number of particles due to finite system effects.  Given the
dimensions of the box and the number of particles, we can then assign the
appropriate wave vectors and $\chi$ value to the system. The potential energy of
the system is then quenched using the conjugate gradient method followed by a
quenching using the MINOP algorithm as was detailed above. 

To characterize inherent structures, we use several quantitative metrics. The
structure factor $S(k)$, {\it cf.} Eq.\ (\ref{eq:sk}), provides structural
information on the long-ranged ordering of the system.  Defined above,
$\varepsilon$ is a scaled potential energy and $\eta$ is a metric for the extent
of stealthiness. In addition, we use the small-$k$ features of $S(k)$ to compare
the extent of hyperuniformity between structures.  Hyperuniform systems have the
feature that $S(k)$ vanishes as $k$ approaches zero. We fit the small-$k$ regime
({\it i.e.} $k<0.5$) to a log-polynomial equation
\begin{equation}
\label{eq:logpoly}
\log S(k) = C_0 + C_1k + C_2k^2 + C_3k^3.
\end{equation}
The parameter $C_0$ can be considered to be a metric for the extent of
hyperuniformity, where a value that diverges to $-\infty$ indicates that a
system is hyperuniform. This metric is best used for relative comparisons among
systems since numerical precision is limited. For example, one could say system
A is more hyperuniform than system B because it has a lower value of $C_0$. 
However, $C_0$ is not be a sufficient test to determine if a configuration is
absolutely hyperuniform ({\it i.e.,} one should refrain from saying system A is
absolutely hyperuniform).

We also employ the bond-order parameter $\Psi_6$ defined as 
\begin{equation}
\Psi_6 = \left| \frac{1}{N_{bonds}}\sum_j\sum_k e^{6i\theta_{jk}}\right|
\end{equation}  
where $\theta_{ij}$ is the angle between two particles with respect to a fixed,
but arbitrary, coordinate axis, to quantify the local orientation order.
Particles are considered ``bonded'' if $r_{ij} < 10$, chosen so as to include
nearest-neighbors. For the perfect triangular lattice, $\Psi_6$ has a value of
unity while for the ideal gas $\Psi_6$ vanishes. Wavy crystals have value of
$\Psi_6$ that can range from about 0.4 to 0.8.

We introduce a configurational proximity parameter $p$ defined as 
\begin{equation}
 p = \frac{ \left(\sum_{i=1}^N|\br_{o,i}-\br_{f,i}|^2\right)^{1/2}    }{r_{NN}}, 
\end{equation}
where $\br_{o,i}$ and $\br_{f,i}$ are the positions of the RSA configuration and
ground state configuration respectively and $r_{NN}$ is the
mean-nearest-neighbor in the RSA configuration. This metric $p$ quantifies the
closeness of two different particle configurations. In what follows, we will
apply it to characterize the relative particle displacements required for a
configuration to relax to the ground-state energy.

\section{Results: Cooling and Quenching}
\label{sec:cooling}

Before discussing the results of the quenching and cooling simulations, we
briefly discuss some aspects of system-size effects related to the inherent
structures of this system.  While we previously showed that system-size effects
were negligible when characterizing the ground states\cite{batten2008cdg} and
the equilibrium properties,\cite{batten2009interactions} we find that there are
appreciable system-size effects when considering the inherent structures and
energy landscape.

We have compared inherent structures consisting 418 particles to 2340 particles
sampled from the same temperature.  In general, larger systems provide a
slightly deeper quenching of the potential energy. In addition, the bond-order
parameters for larger systems tend to be less than that of smaller systems. This
is expected since larger systems often have various polycrystalline-like domains
that impart destructive interference on $\Psi_6$.  The fraction of quenchings
from a fixed temperature that achieve the ground state varied considerably
across system sizes.  While the quenching of the energy of small systems
occasionally resulted in achieving a ground-state structure, quenches of the
large systems never resulted in a ground-state structure.  The measure of the
ground-state manifold becomes increasingly small as the system size increases.
Given that the number of inherent structures increases exponentially with
respect to system size,\cite{stillinger1999exponential} it is unsurprising that
larger systems would not sample the capture basins for ground states as
frequently as smaller systems. System-size effects have been also been
identified in the binary Lennard-Jones system.\cite{buchner1999potential}  Due
to the system-size effects, we should not extrapolate true dynamical information
from our results.  Here, we present the analysis for 418-particle systems,
noting that the trends across $\chi$ values are similar to those in the
2340-particle system.

The properties of the excited-state (or thermalized) structures behave as
expected.  The thermodynamic properties lagged the equilibrium values as the
system was cooled, and for slower cooling rates, the lag in thermodynamic
properties was smaller than that for fast cooling rates. However, we find that
the rate at which the system was cooled provided no distinguishable effects on
the properties of the inherent structures when compared across various cooling
rates.  This should be contrasted with conventional glass-forming systems such
as binary Lennard-Jones where systems cooled under slow rates found deeper wells
in the energy landscape than systems cooled under faster
rates.\cite{sastry1998signatures}

In our previous research, we have identified the approximate range for
transitions from crystalline or wavy-crystalline to liquids to be in the
temperature range of $T^*<0.00075$.\cite{batten2009interactions} We find that
the temperature at which the inherent structures were sampled from had no
distinguishable effect on the thermodynamics properties or structures. As the
system was cooled, the systems hopped across various capture basins without
regard to temperature.

\begin{figure}
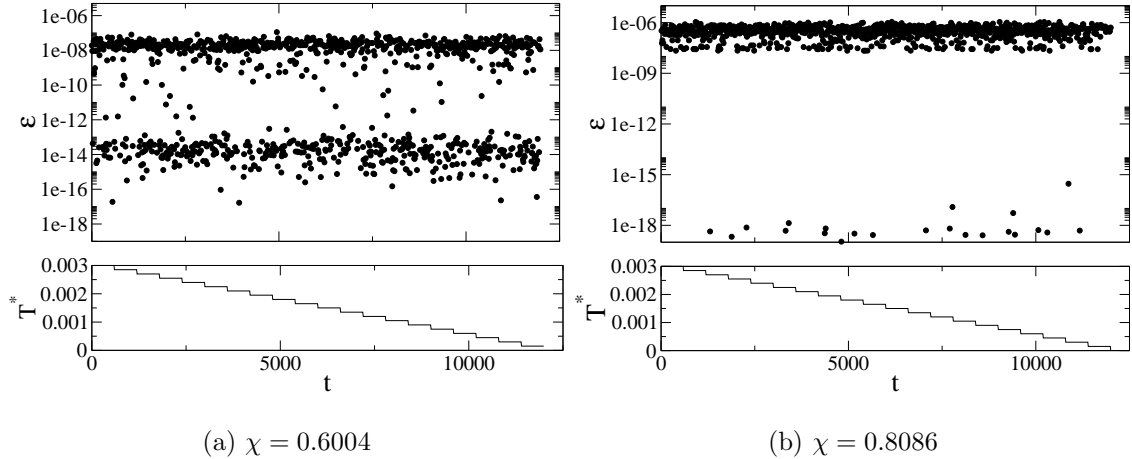

\subfigure[ $\text{ }\chi =
0.6004$]{\label{fig:pottime0.6}\includegraphics[width=0.45\textwidth,
clip=true]{Fig3a.eps}}
\subfigure[ $\text{ }\chi =
0.8086$]{\label{fig:pottime0.8}\includegraphics[width=0.45\textwidth,
clip=true]{Fig3b.eps}}
\caption{ Scaled potential energy $\varepsilon$ of inherent structures as a
function of time for $\chi$=0.6004 and 0.8086.  The lower plots show the
temperature schedules as a function of time.  The temperature was reduced in a
piecewise linear manner with an effective cooling rate of $\gamma = 10^{-7}$.
Inherent structures were sampled every 30 time steps. As a function of time, the
systems continually hop from one capture basin to another since no two
consecutively-sampled structures have identical energies. We found that the
effective cooling rate did not have a significant effect on the fluctuations of
the energies of inherent structures. At low temperatures, systems evidently are
not trapped in deep energy wells.  For $\chi=0.6004$, a significant fraction of
inherent structures have an energy below the ground-state threshold of
$\varepsilon < 10^{-10}$, while only a few inherent structures lie below this
threshold for $\chi=0.8086$.}
\label{fig:pottime2}
\end{figure}

Figure \ref{fig:pottime2} exhibits the dynamics of systems as they sample
inherent structures for $\chi$=0.6004 and 0.8086 for 418 particles cooled at a
rate of $\gamma$=$10^{-7}$. The figure shows the potential energy of inherent
structures sampled every 30 time steps. In the figure, there are two classes of
behaviors. For $\chi=0.6004$, the system samples higher-energy inherent
structures and also frequently samples ground states, those systems with
$\varepsilon < 10^{-10}$.  For $\chi=0.8086$, the system samples mostly
higher-energy structures and occasionally, though far less frequently than for
smaller $\chi$, samples ground states.  For both values of $\chi$, the system
hops to a new inherent structure between samplings. There were no consecutive
samples that were identical except for very low temperatures far below the
melting point ({\it e.g.,} $T^*=0.00015$). The rate at which the system samples
capture basins does not appear to slow down significantly when the temperature
is reduced as one might expect.  This phenomenon occurred for all cooling rates
we tested.  In addition, even when sampling inherent structures every three time
steps, the system never remained in the same basin for consecutive samplings.
Note that the depths of the quenches for ground-state structures were deeper for
$\chi=0.8086$ than for $\chi=0.6004$.  This demonstrates the issue that arises
with finite-precision computing because these scaled energies should vanish to
zero in a mathematical sense.  In a practical sense, systems with energies below
this threshold and orders of magnitude below the clusters of systems higher up
are ground states. 

\begin{figure}
\includegraphics[width=0.6\textwidth, clip=true]{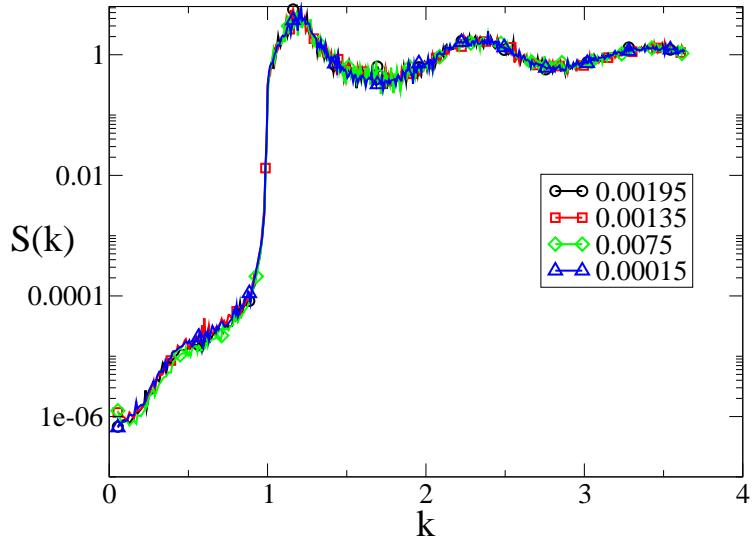}
\caption{(Color online) Structure factor for inherent structures with
$\chi$=0.6004 and $N$=418 and at various dimensionless temperatures $T^*$ -
0.00195 (circles), 0.00135 (squares), 0.0075 (diamonds), and 0.00015
(triangles). Each temperature is ensemble averaged over 50 configurations
sampled from a molecular dynamics trajectory. Despite the range of temperatures,
temperature has little effect on the inherent structures.}
\label{fig:skavgtemp}
\end{figure}

We find that the structural features of the inherent structures did not vary
with temperature either. Figure \ref{fig:skavgtemp} shows the structure factor
for inherent structures with $\chi=0.6004$ ensemble-averaged for several
temperatures. However, when observing the structure factors for individual
configurations, there is some small variability in the shape of $S(k)$. However,
there are no clear correlations with temperature.  In the figure, the structure
factors associated with each temperature overlay nearly perfectly. Even at very
low temperatures ($T^*=0.00015$), the structure factor for the inherent
structures does not differ from those obtained from a liquid state.

For familiar systems, as the temperature is reduced, the potential energy of the
inherent structures typically becomes lower as the system continues to reduce
the potential energy through structural rearrangements, sampling deeper and
deeper basins. However, we find as the temperature is reduced, the energy of the
inherent structures does not appear to decrease. This reveals that the energy
landscape is relatively flat and insensitive to temperature. The heights of the
barriers, which we explore in Sec.\ \ref{sec:mapping}, are therefore expected to
be relatively small.

\begin{figure}
\includegraphics[width=0.6\textwidth, clip=true]{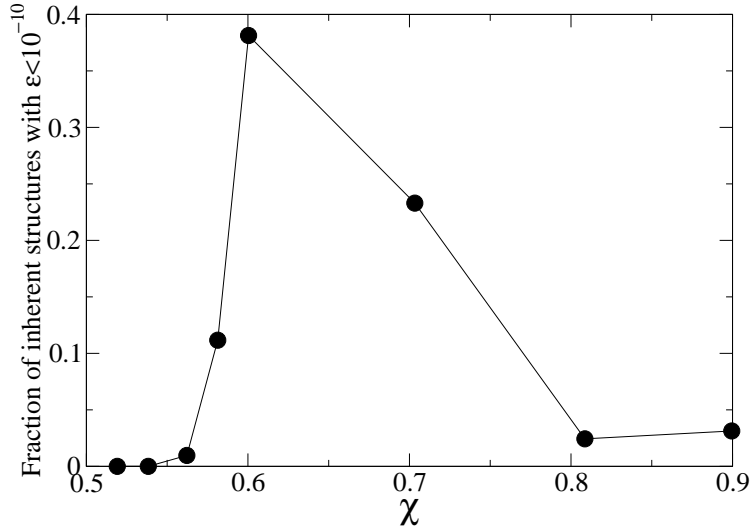}
\caption{Fraction of inherent structures obtained via the cooling and quenching
procedure with $\gamma=10^{-8}$ and $N=418$ that are ground states ({\it i.e.,}
scaled potential energy $\varepsilon<10^{-10}$).  Data represent the average of
four independent cooling runs with 750 samples for each run.  $\chi$ values of
0.6004 and 0.7033 achieved the highest fraction of ground states. }
\label{fig:gsfrac}
\end{figure}

In Fig.\ \ref{fig:gsfrac}, we plot the frequency in which the cooling and
quenching procedure yielded a ground state for 418 particles.  The plot shows
the fraction of inherent structures that were ground states as a function of
$\chi$ for 3000 inherent structures for each $\chi$. We have found that the
temperature from which the inherent structures are sampled does not have a
significant effect on the rate at which the ground state is achieved. However,
the value of $\chi$ affects the frequency at which the ground state is found. 
In general, as $\chi$ is increased from 0.6 to 0.9, the rate at which the ground
states are found is reduced. However, for the ``disordered-ground-state''
region, $\chi < 0.58$, ground states were found infrequently. For example, in
Fig.\ \ref{fig:gsfrac}, the frequency in which ground states are found for
$\chi=0.5622$ is much smaller than that of $\chi=0.6004$.  This suggests that
the relative availability of the ground-state manifold is small for $\chi<0.58$,
becomes larger in the wavy-crystalline region ($\chi>0.58$), and then diminishes
as the terminal $\chi$ of 0.9 is reached.  This shows that changes in $\chi$
(and equivalently, density for fixed $K$), fundamentally affect the shape and
size of the ground state manifolds. An expansion or compression of the system
does not simply rescale the energy landscape while maintaining its overall
shape.  Systems containing 2340 particles never achieved ground state energies,
presumably because of the exponential growth in the number of inherent
structures with system size.\cite{stillinger1999exponential} 

\section{Results: Structural Characteristics}
\label{sec:structure}

\begin{figure}
\includegraphics[width=0.3\textwidth, clip=true]{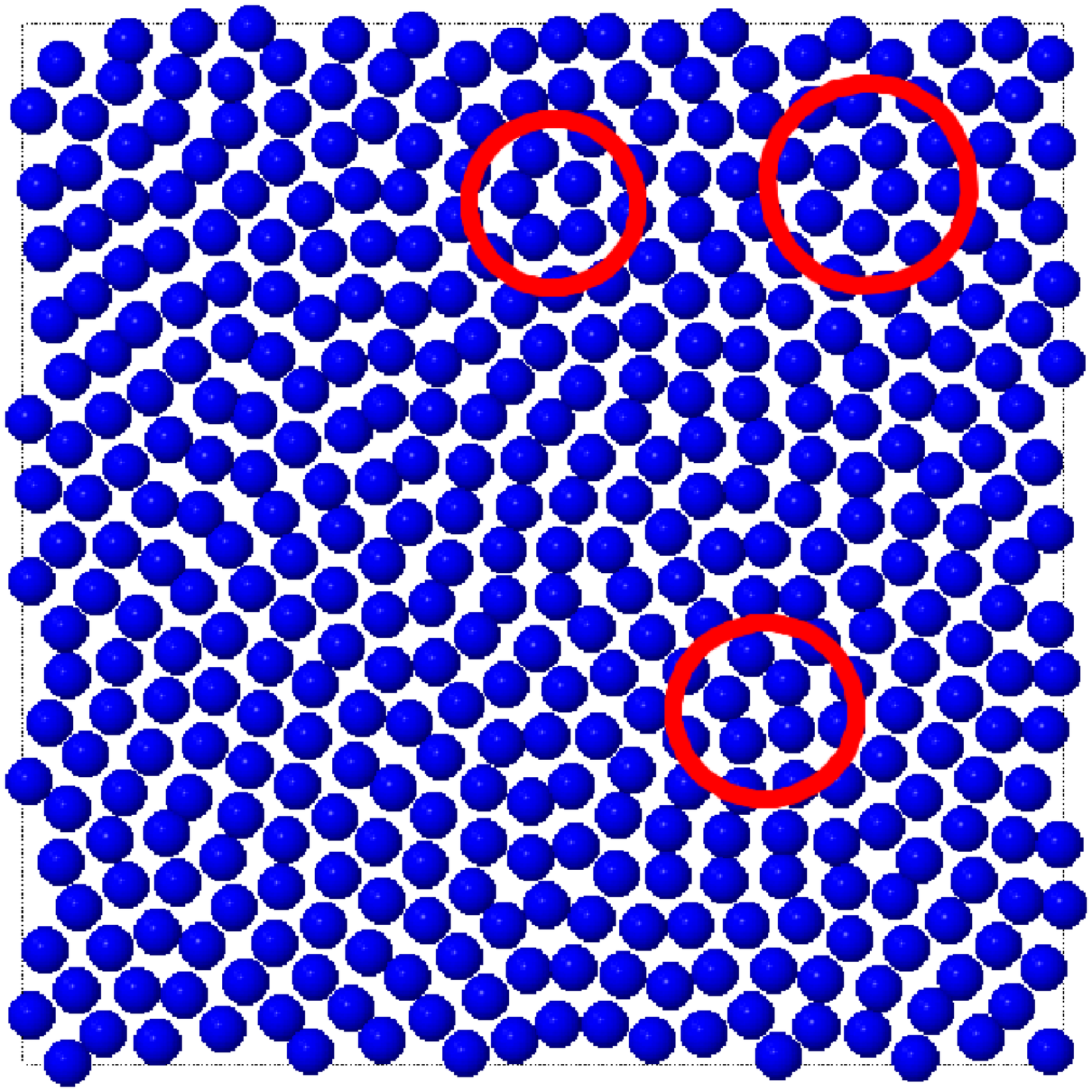}
\includegraphics[width=0.3\textwidth, clip=true]{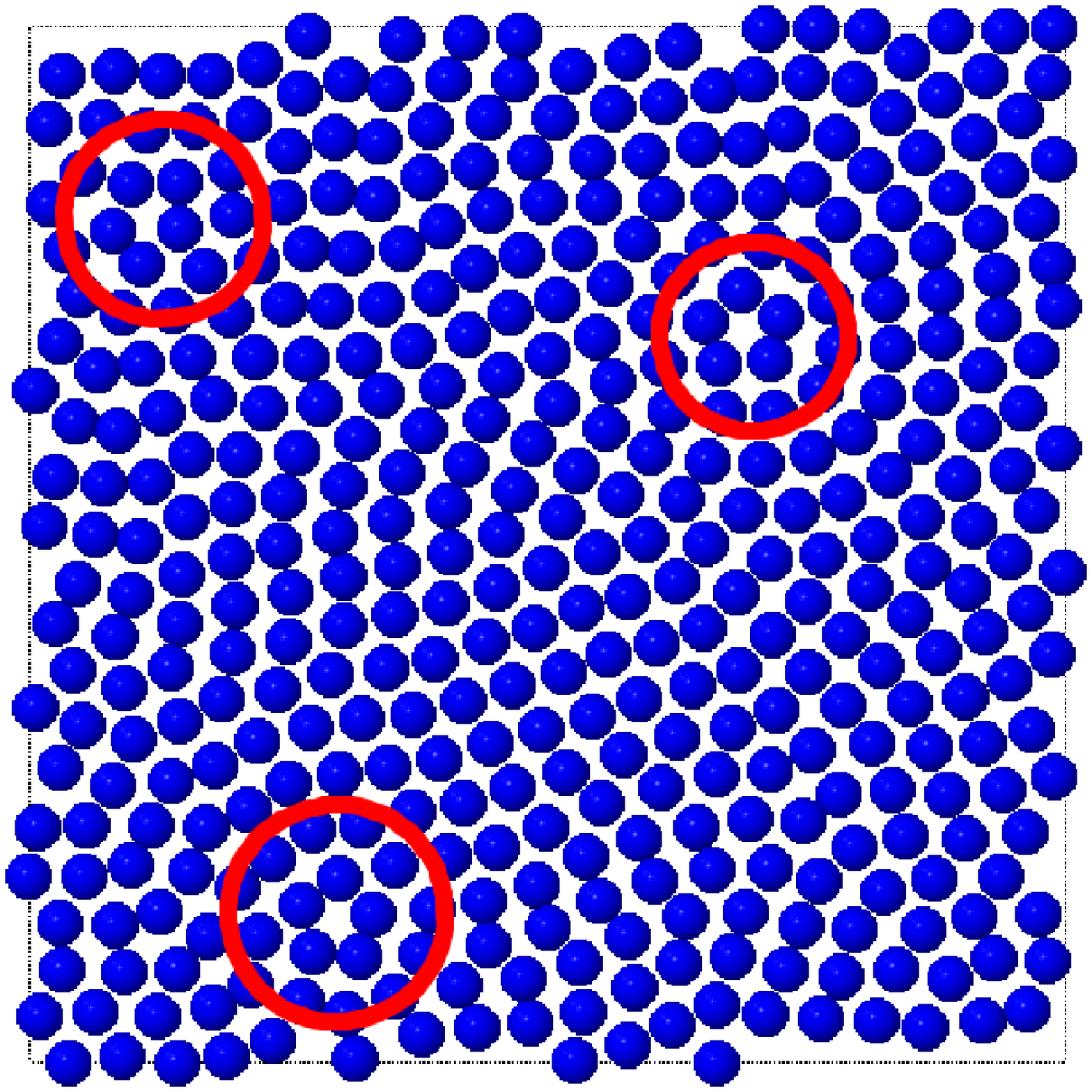}
\includegraphics[width=0.305\textwidth, clip=true]{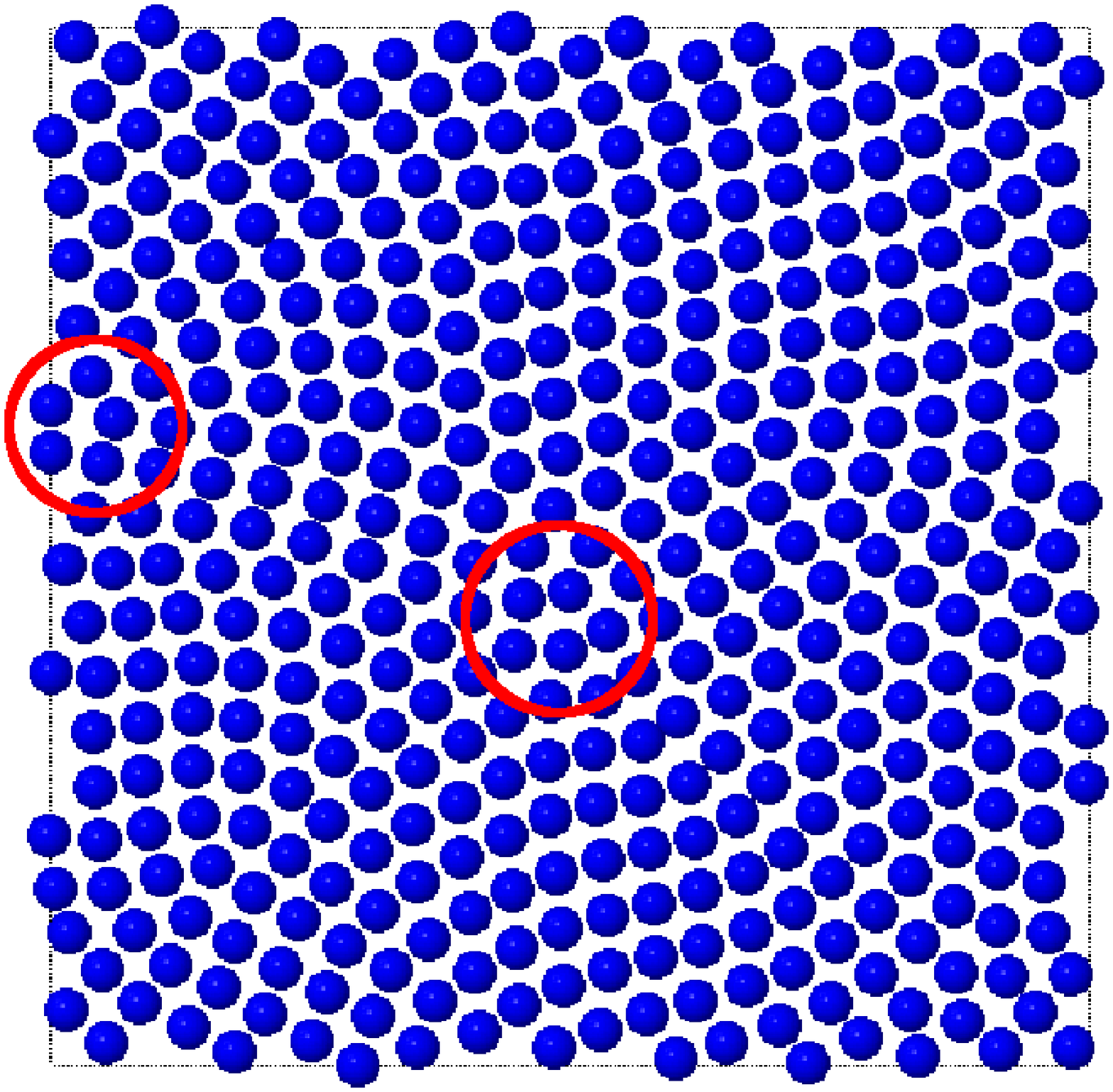}
\caption{(Color online) Inherent structures produced from the cooling and
quenching procedure sampled from a high-temperature liquid ($T^*=0.0018$) for
$\chi= 0.5191$ (left), 0.5622 (middle), and 0.6004 (right) for 418 particles.
These structures demonstrate the five-particle clusters that are common in
inherent structures for these $\chi$ values.  The five-particle rings are
highlighted by a surrounding circle. The buildup of wavy grains are visible for
$\chi=0.6004$.  }
\label{fig:inhsmall}
\end{figure}

\begin{figure}
\includegraphics[width=0.3\textwidth, clip=true]{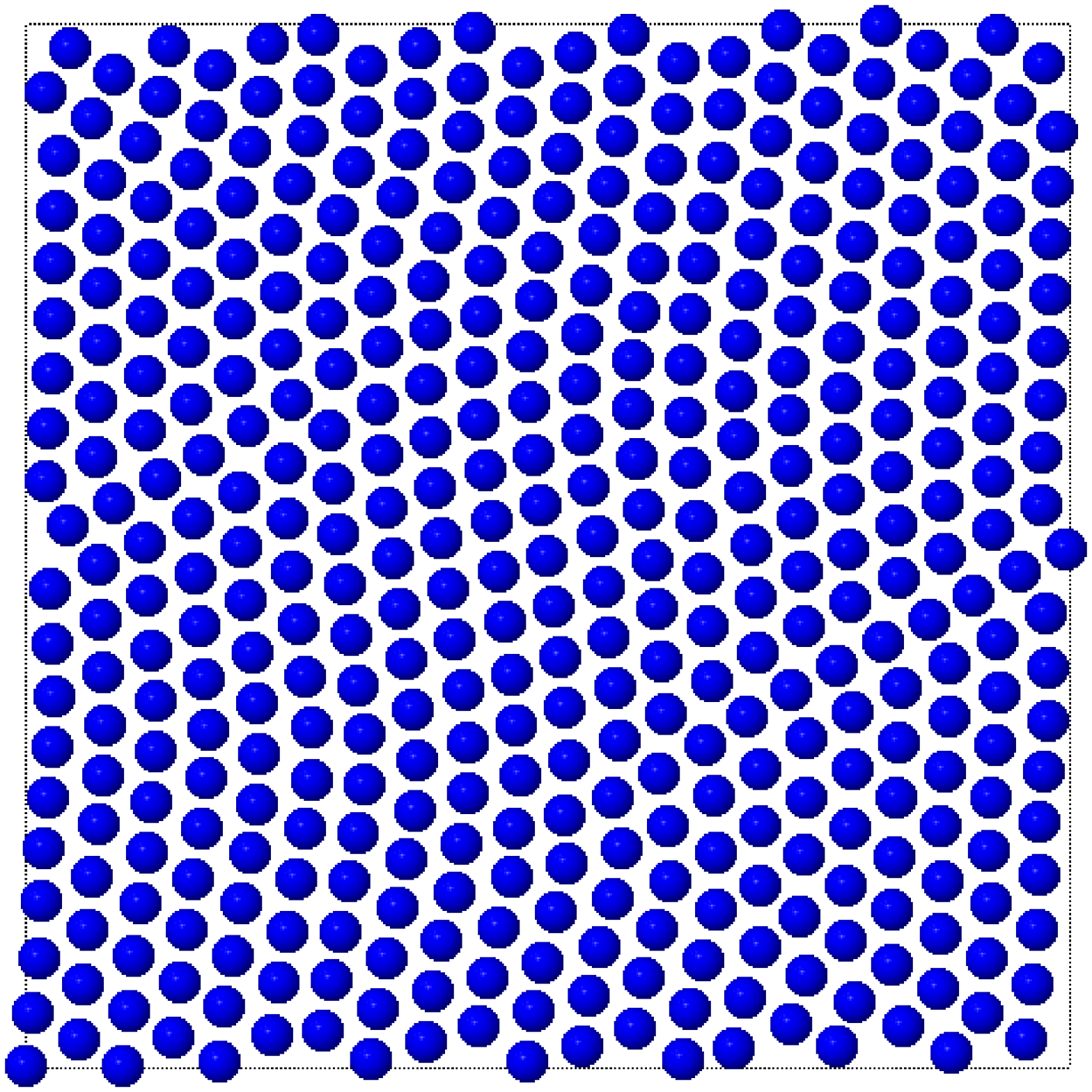}
\includegraphics[width=0.3\textwidth, clip=true]{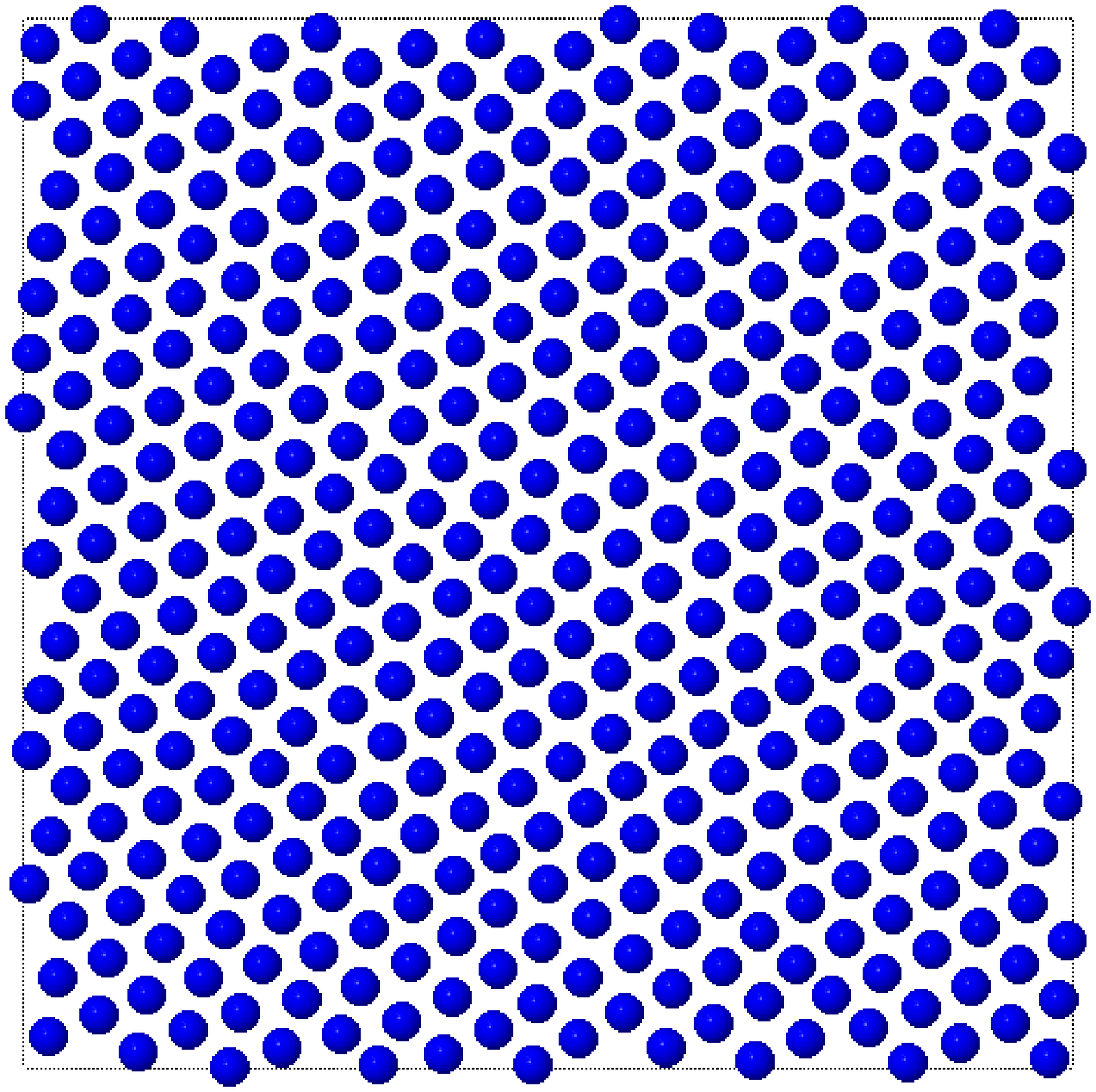}
\includegraphics[width=0.3\textwidth, clip=true]{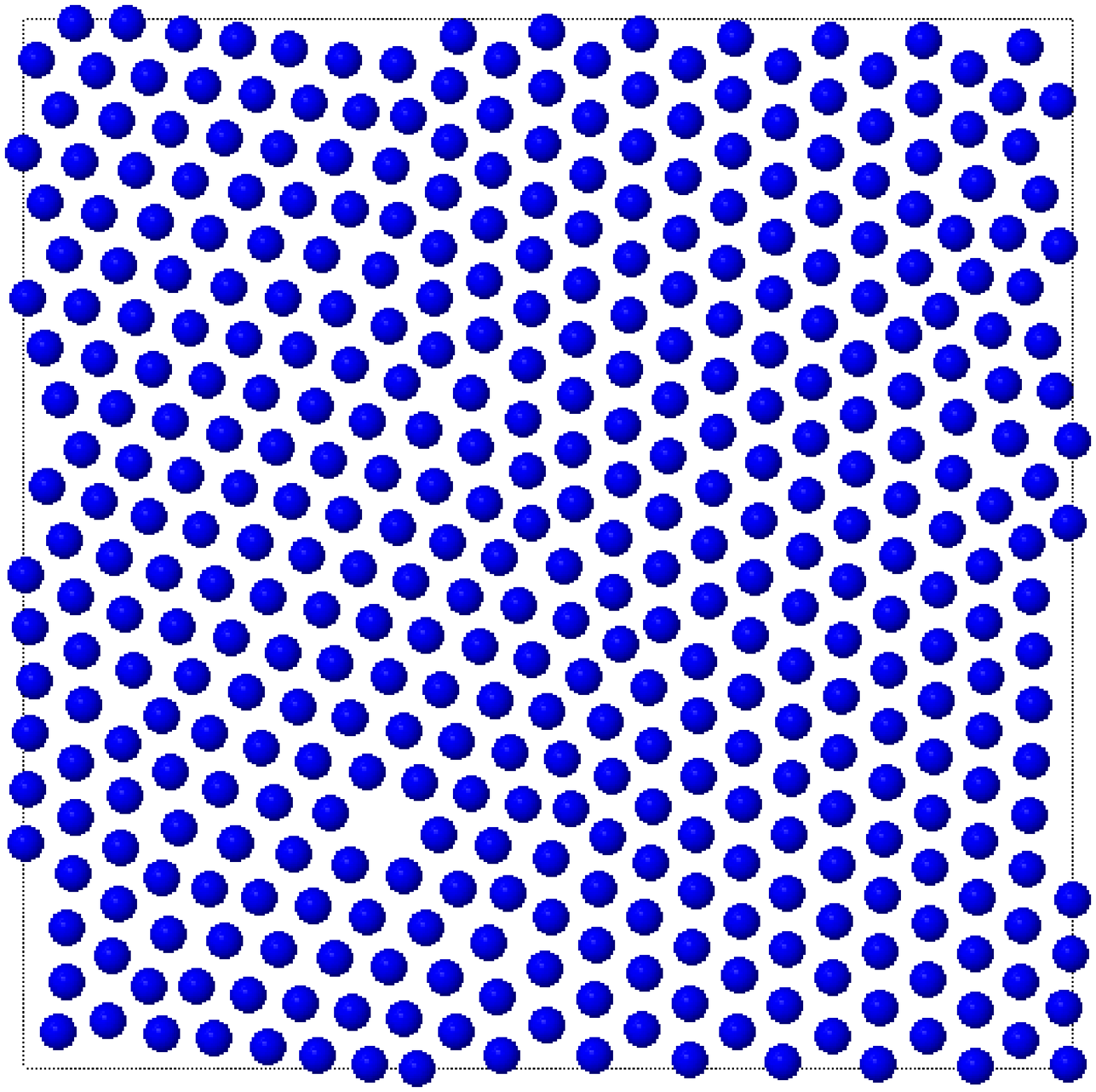}
\caption{(Color online) Inherent structures produced from the cooling and
quenching procedure sampled from a high-temperature liquid ($T^*=0.0018$) for
$\chi= 0.7033$ (left), 0.8086 (middle), and 0.8995 (right) for 418 particles.
These structures demonstrate the wavy grains and vacancy-interstitial defects
common to inherent structures for $\chi > 0.58$. As $\chi$ increases, the
inherent structures are more ordered, though grain boundaries prevent the
systems from minimizing the energy.}
\label{fig:inhlarge}
\end{figure}

Figures \ref{fig:inhsmall} and \ref{fig:inhlarge} show inherent structures
obtained via the cooling and quenching procedure. These structures display
features that are common among inherent structures. Upon examining the inherent
structures, we found several structural motifs - five-particle rings, wavy grain
boundaries, and vacancy-interstitial defects. The frequency of these features
varies depending on $\chi$.  For $0.5 < \chi < 0.6$, the five-particle ring, as
demonstrated in Fig.\ \ref{fig:inhsmall}, is the most prevalent feature.  The
five-particle ring arises due to the local interactions of constituent
particles. The nearest- and second-nearest neighbor distances lie within the
first and second wells of $v(r)$ respectively. The ratio of the relative
distances creates a frustration that favors the local five-particle ring.
Evidently, these local forces are more dominant than the long-range forces that
promote stealthiness.

For $0.58 < \chi < 0.89$, a common feature of inherent structures is a wavy
grain boundary as evident in Figs.\ \ref{fig:inhsmall} and \ref{fig:inhlarge}.
In these images, the grain boundaries manifest themselves in a more wavy form
than one might see in a Lennard-Jones inherent structure, in which particles
prefer to align in straight lines. This $\chi$ range is associated with
wavy-crystalline ground states, which appear as a uniformly sheared triangular
or square lattice.  This behavior evidently manifests itself in the inherent
structures where the grain boundaries appear to have some waviness.  For these
$\chi$ values, the long-range nature of the potential is sufficiently strong to
inhibit five-particle rings in favor of the wavy grains.  Lastly, for
$\chi=0.8995$, the last image in Fig.\ \ref{fig:inhlarge}, the inherent
structures either have a large grain boundary of perfectly aligned particles,
vacancy-interstitial pairs, or both.  These inherent structures also exhibit
polycrystallinity. In the cases of $\chi>0.58$, the multiple wavy domains appear
to be in structural conflict with each other. This is best demonstrated by Fig.\
\ref{fig:inhsmall} for $\chi=0.6004$ where several domains appear to meet in the
middle of the system.  This conflict evidently ``jams'' the system and prevents
it from reaching the ground state. 

\begin{figure}
\includegraphics[width=0.6\textwidth, clip=true]{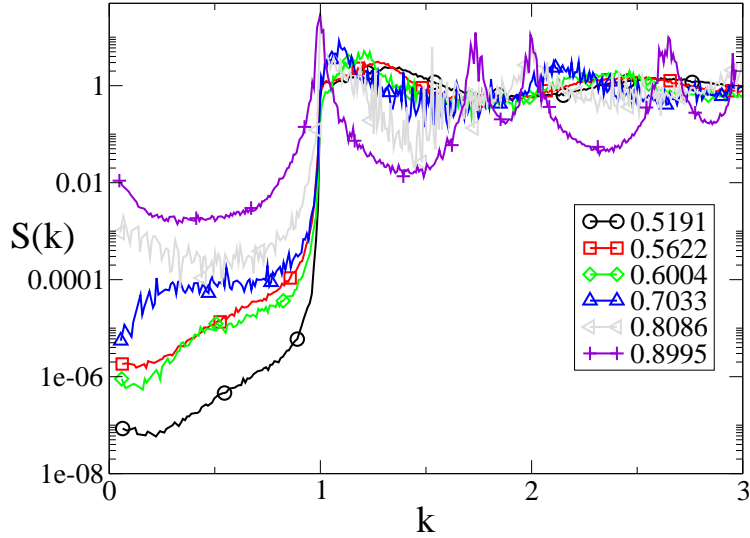}
\caption{(Color online) Structure factor for inherent structures as a function
of $\chi$ ensemble averaged over 50 configurations sampled from a molecular
dynamics trajectory at $T^*=0.00105$ for a system of 418 particles.  As $\chi$
increases, the inherent structures become less stealthy. Also, the extent of
hyperuniformity decreases when $\chi$ is increased. }
\label{fig:skavginh}
\end{figure}

Observing the structure factor for inherent structures shows that inherent
structures with smaller $\chi$ are more stealthy and more hyperuniform than
those at higher $\chi$. However, inherent structures that are not ground states
were neither completely stealthy nor hyperuniform. In Fig.\ \ref{fig:skavginh},
we show the ensemble-averaged $S(k)$ for 50 inherent structures sampled at
$T^*=0.00105$ for various $\chi$.  For the smallest $\chi$ value, the small-$k$
tail of the structure factor initially decreases and then rapidly increases as
$k$ is increased toward unity.  The shape of the small-$k$ tail of $S(k)$
remains constant but is shifted to higher values (less hyperuniform) as $\chi$
increases to $\chi=0.5813$.

However, as $\chi$ leaves the disordered-ground-state regime, the shape of the
curve changes, as exemplified by the curves for $\chi$=0.6004 and 0.7033. In
Fig.\ \ref{fig:skavginh}, the curve for $\chi=0.6004$ demonstrates the new
shape. For the small-$k$ region, $S(k)$ initially has a negative slope which
quickly turns positive, and nearly flattens before diverging near $K$. The shape
of this ensemble-averaged $S(k)$ is due to the relatively large number of ground
states that arise in the inherent structure analysis for this $\chi$. For
$\chi=0.6004$, nearly 40\% of inherent structures are ground states, and this
$S(k)$ represents the average of the ground-state $S(k)$ and the inherent
structure $S(k)$.  A similarly large fraction of inherent structures are ground
states for $\chi$=0.7033. If the ground states are removed from the
ensemble-averaging, the structure factor appears more like those associated with
$\chi$=0.8086 and 0.8995. For these $\chi$, $S(k)$ has a initial negative slope
that turns positive near $k$=0.5, which then diverges near $K$.

It is interesting to note that the way in which systems get ``stuck'' in an
inherent structure for $\chi$ in the disordered regime is different than that of
the wavy-crystalline regime.  For $\chi<0.58$, the systems struggle most to
minimize $S(k)$ for $k>0.5$. The length scale associated with $k>0.5$
corresponds to local interactions in real space ({\it i.e.,} on the order of a
few particle diameters). This manifests itself with the five-particle rings. For
$\chi>0.58$, the systems have difficulty minimizing $S(k)$ near zero and unity.
The $k$-values near zero represent longer-range interactions than the $k$-values
near unity. The inherent structures for $\chi>0.58$ are therefore ``stuck'' on a
longer length scale. It is no surprise then that the inherent structures appear
locally ordered but are frustrated with a grain boundary.

\begin{figure}
\includegraphics[width=0.6\textwidth, clip=true]{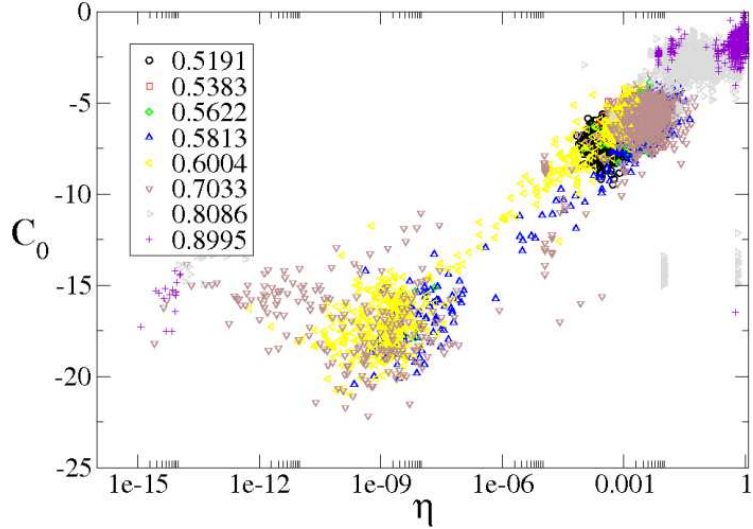}
\caption{(Color online) Hyperuniformity parameter $C_0$ versus
stealthiness metric $\eta$ for inherent structures generated from the MD and
quench procedure. There is a clear relation between the hyperuniformity
coefficient $C_0$ and the stealthiness metric $\eta$, although there exist
outlying configurations. There is also a positive relation between $C_0$ and
$\chi$. Those configurations in the lower left with $\eta < 10^{-9}$ are
typically ground states.  800 configurations are shown for each $\chi$ value.}
\label{fig:c0stealth}
\end{figure}

It is obvious that as $\chi$ increases the extent of hyperuniformity of the
inherent structures diminishes.  This is shown in an ensemble sense in Fig.\
\ref{fig:skavginh}. However, we have also shown this general relationship for
individual structures. In Fig.\ \ref{fig:c0stealth}, we display the
hyperuniformity coefficient $C_0$ for the log-polynomial fit to $S(k)$ in Eq.\
(\ref{eq:logpoly}). In the figure, each data point represents one inherent
structure. There are two clusters of points, where the cluster of points on the
lower left are typically ground states where $S(k)$ vanishes as $k$ approaches
zero, but due to numerical precision the hyperuniformity coefficient $C_0$
remains finite.  In the cluster in the upper right of Fig.\ \ref{fig:c0stealth},
there is a linear relation between $C_0$ and $\log \eta$. Also, it is clear that
as $\chi$ increases, the hyperuniformity parameter $C_0$ also increases.  A few
outlying structures are present for $\chi$=0.7003, 0.8086 and 0.8995. These
structures appear as triangular lattices with very slight waves or shear applied
to them, a sufficient amount that they are not ground states, but still have
very small limiting values of $S(k)$ as $k$ approaches zero.

\begin{figure}
\includegraphics[width=0.6\textwidth, clip=true]{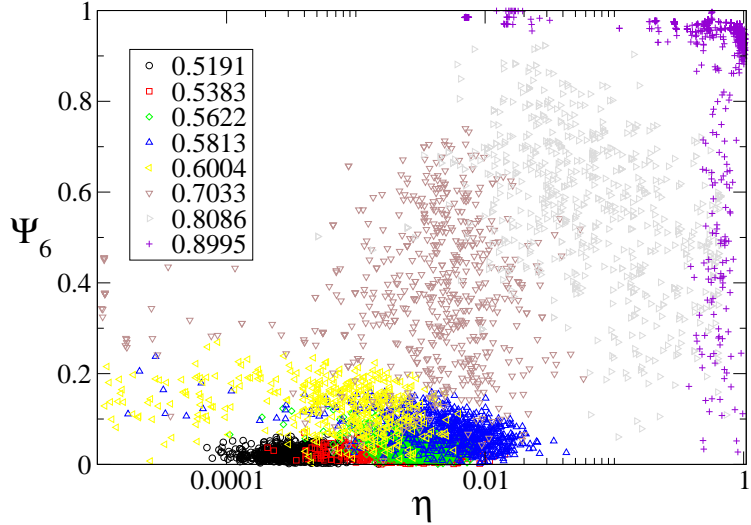}
\caption{(Color online) Bond-order parameter $\Psi_6$ versus stealthiness metric
$\eta$ for inherent structures generated from the MD and quench procedure. As
$\chi$ increases, there is an increasing degree of six-fold ordering. However,
for large $\chi$, the range of available $\Psi_6$ extends larger due to the
polycrystalline nature of the configurations. 800 configurations are shown for
each $\chi$ value. For clarity, we omit those structures with $\eta<10^{-5}$ so
that the plot can provide more detail about the inherent structures. }
\label{fig:potbond2}
\end{figure}

The stealthiness of the inherent structures, as characterized by the metric
$\eta$, in general does not have a relation to the six-fold bond-order
orientiational order of a system. Figure \ref{fig:potbond2} shows the bond-order
parameter $\Psi_6$ as a function of stealthiness $\eta$ for various inherent
structures.  For each $\chi$ value, there is a range of available $\eta$ and
$\Psi_6$. Initially, for $\chi$ near 0.5, these ranges are rather narrow, and as
$\chi$ increases, these ranges become much broader. For large $\chi$=0.8086
there is a large diversity of $\Psi_6$ and $\eta$ available for inherent
structures. However, for $\chi$=0.8995, there is a narrow band of available
$\eta$ but a large range of $\Psi_6$ depending on the number and orientation of
opposing crystalline domains. The group of structures for $\chi$=0.8995 and
$\eta\sim0.02$ are those structures with only vacancy-interstitial defects.

\begin{figure}
\includegraphics[width=0.6\textwidth, clip=true]{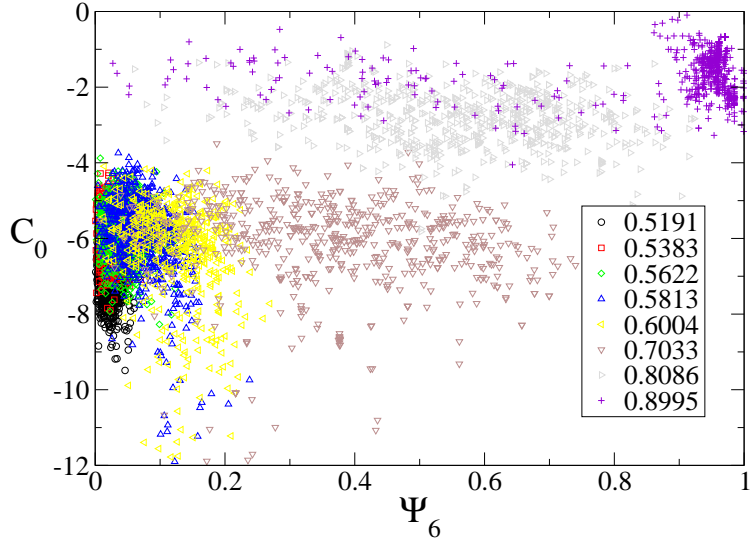}
\caption{(Color online) Hyperuniformity parameter $C_0$ versus bond-order
parameter $\Psi_6$ for inherent structures generated from the MD and quench
procedure.  Despite the broad range of six-fold ordering at larger $\chi$, the
level of hyperuniformity remains consistent across $\chi$ values. 800
configurations are shown for each $\chi$ value.  For clarity, we omit those
structures with $C_0<-12$ so that the plot can provide more detail on the
inherent structures. }
\label{fig:c0bond}
\end{figure}

The bond-order parameter also shows considerable variation when plotted against
the hyperuniformity parameter $C_0$.  Figure \ref{fig:c0bond} shows $C_0$ as a
function of $\Psi_6$ for various $\chi$ values. As $\chi$ increases, the
variation in $\Psi_6$ grows while $C_0$ generally falls into a narrow range of
values.  This figure demonstrates that hyperuniformity does not necessarily
require local six-fold ordering. Systems with $\chi$=0.7033 have the largest
range of available local six-fold ordering for more hyperuniform structures
because $\Psi_6$ ranges from as low as 0.1 to just below unity.

\section{Paths from Inherent Structures to Ground States}
\label{sec:mapping}

\begin{figure}
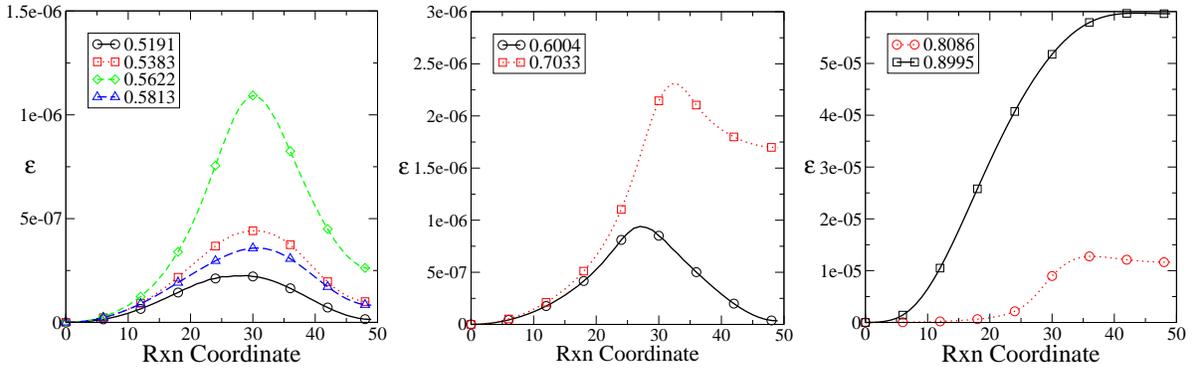

\includegraphics[width=0.31\textwidth, clip=true]{Fig12a.eps}
\includegraphics[width=0.32\textwidth, clip=true]{Fig12b.eps}
\includegraphics[width=0.31\textwidth, clip=true]{Fig12c.eps}
\caption{(Color online) Representative minimum-energy paths from ground states
(point zero on the reaction coordinate) to inherent structures (last point on
the reaction coordinate) along a generalized reaction pathway for $N=418$ for
(a) $0.5<\chi<0.6$, (b) $\chi=0.6004$ and 0.7033, and (c) $\chi=0.8086$ and
0.8995. The local maximum corresponds to a saddle point in the energy landscape.
}
\label{fig:barriers}
\end{figure}

The nudged-elastic-band algorithm was used to determine the minimum-energy path
from a ground-state structure to an inherent structure nearby in configurational
space. We find that the height of the barrier between the ground state and the
inherent structure generally increases with $\chi$. However, there are instances
that deviate from this general situation. Figure \ref{fig:barriers} illustrates
the minimum-energy paths found for various $\chi$ from the ground state (point
zero on the reaction pathway) to the inherent structure (last point on the
reaction pathway). The pathway is discretized over fifty images.  In all the
cases, the landscape is locally quadratic near the ground state and inherent
structure.  The maximum for each pathway represents the saddle point in the
energy landscape separating the ground state and inherent structures.
Connections from the triangular lattice to inherent structures were challenging
to construct because of the zero-energy modes present in triangular-lattice
systems. The nudged-elastic-band algorithm encountered difficulty in traversing
through zero-energy valley and then uphill toward a saddle point. In all cases
except $\chi$=0.8995, we report transitions from non-lattice ground states to
inherent structures.

For $\chi<0.6$ (left in Fig.\ \ref{fig:barriers}), the barrier height,
$\varepsilon_{sad}$ is significantly large compared to the difference in the
energies of the ground-state and inherent structures ({\it i.e.,}
$\varepsilon_{inh} - \varepsilon_{gs} \ll \varepsilon_{sad}$). When $\chi$
becomes larger, as in the case for $\chi$=0.7033 and higher, middle and right in
Fig.\ \ref{fig:barriers}, $\varepsilon_{sad}$ is nearly equivalent in height. 
For the case of $\chi$=0.8995, the minimum-energy path appears to flatten out as
it approaches the inherent structure. In reality, there is a maximum energy
along the path, although it is difficult to discern due to the scaling in the
plot. In the cases where the energy of the inherent structure is nearly equal to
the barrier height, small perturbations from the inherent structure can allow
the system to relax to the ground state. On the other hand, for smaller $\chi$,
relatively higher-energy perturbations are necessary for the inherent structure
to climb the energy barrier and fall to the ground state. While there are
exceptions to this phenomenon, they usually occur when the inherent structure
and ground state are not nearby in the energy landscape ({\it i.e.,} there is a
another inherent structure closer to the ground state). Along these paths, the
configurational proximity metric $p$ monotonically decreases along the path from
the ground state to the inherent structure. The metric $p/N$ for the proximity
per particle between the inherent structure and the ground state is of the order
0.1.  On average, particles are collectively displaced by 10\% of the nearest neighbor
spacing to achieve a ground state.

\begin{figure}
\includegraphics[width=0.3\textwidth, clip=true]{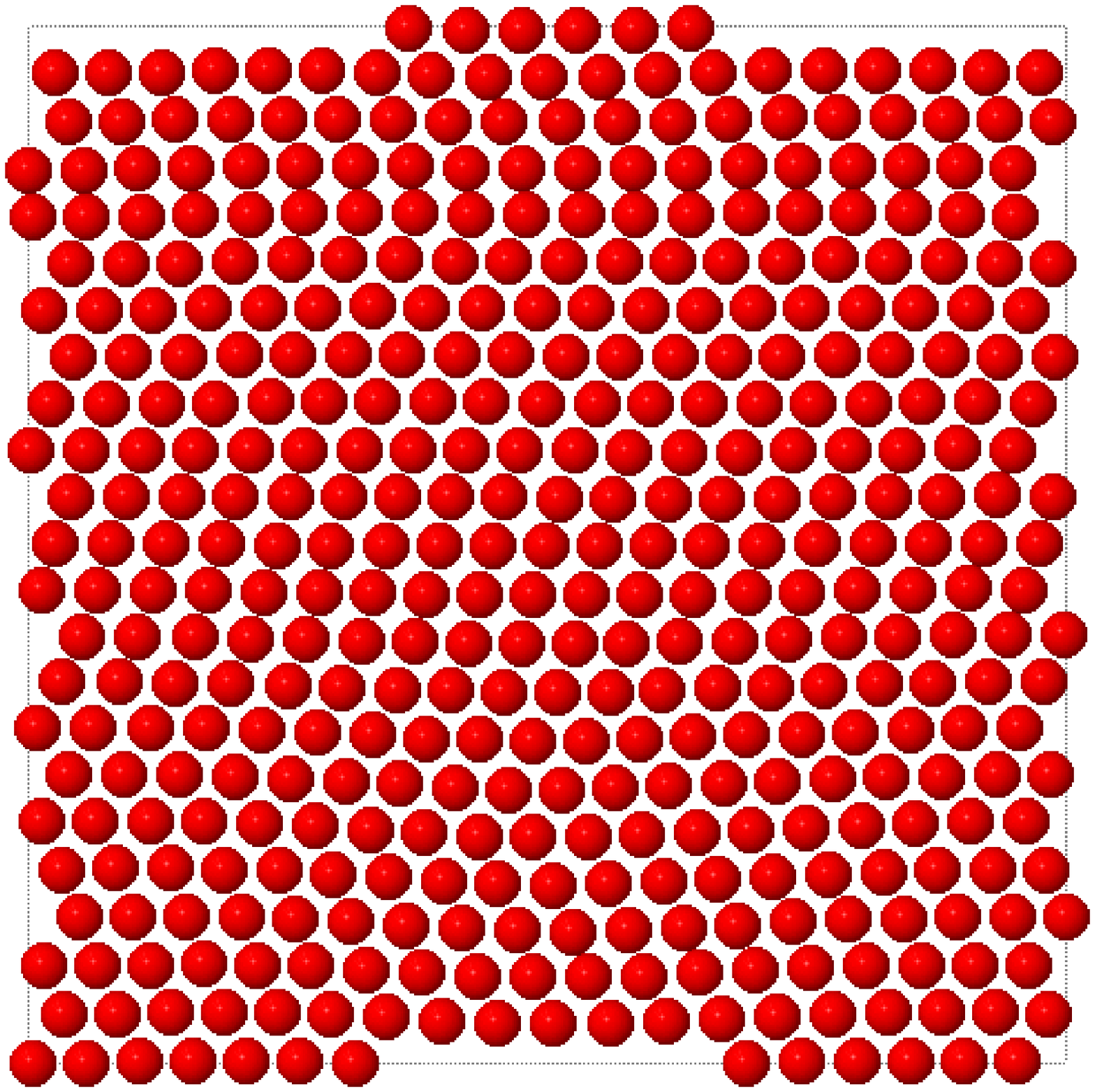}
\includegraphics[width=0.3\textwidth, clip=true]{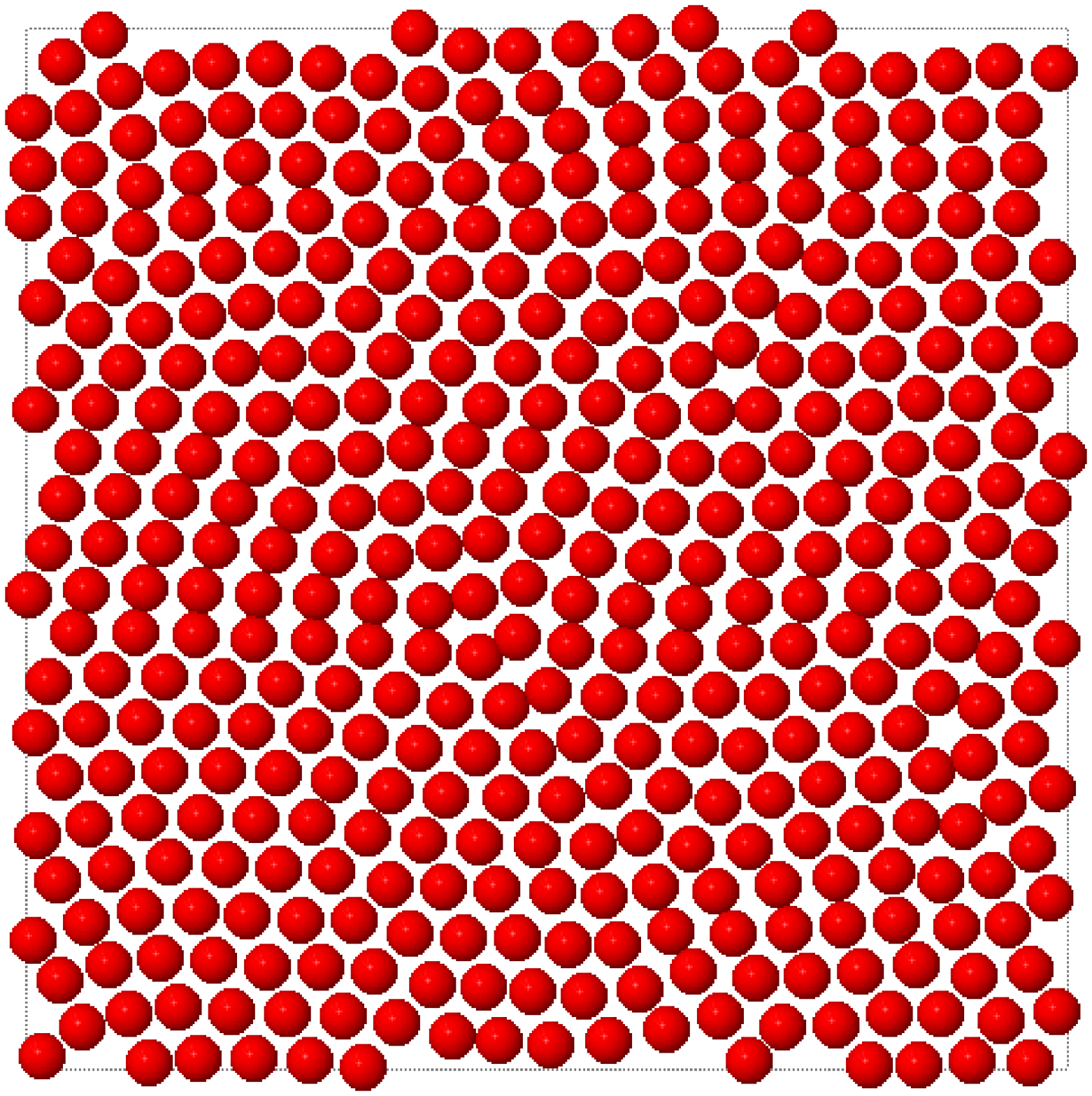}
\includegraphics[width=0.3\textwidth, clip=true]{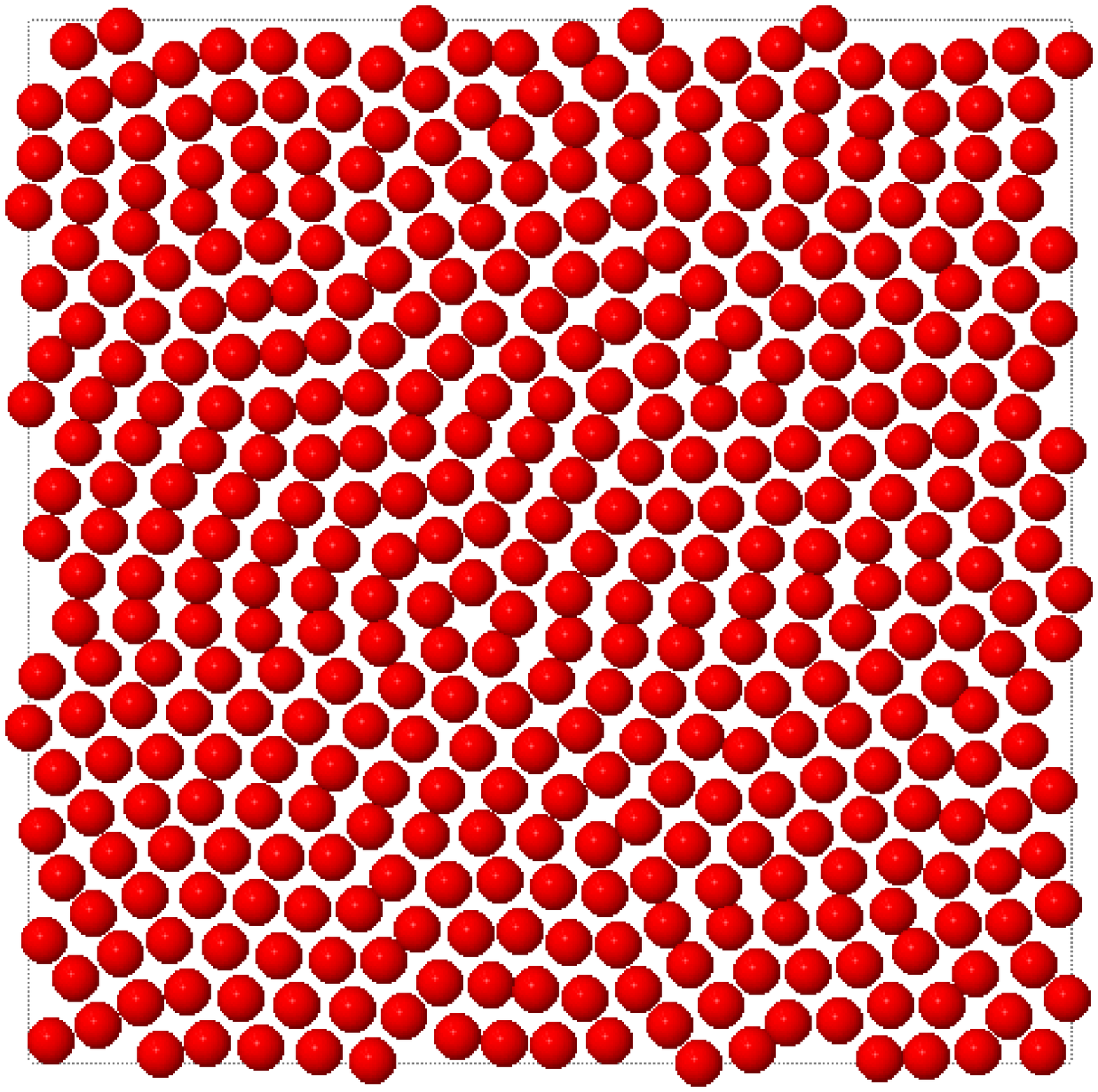}\\
\includegraphics[width=0.3\textwidth,clip=true]{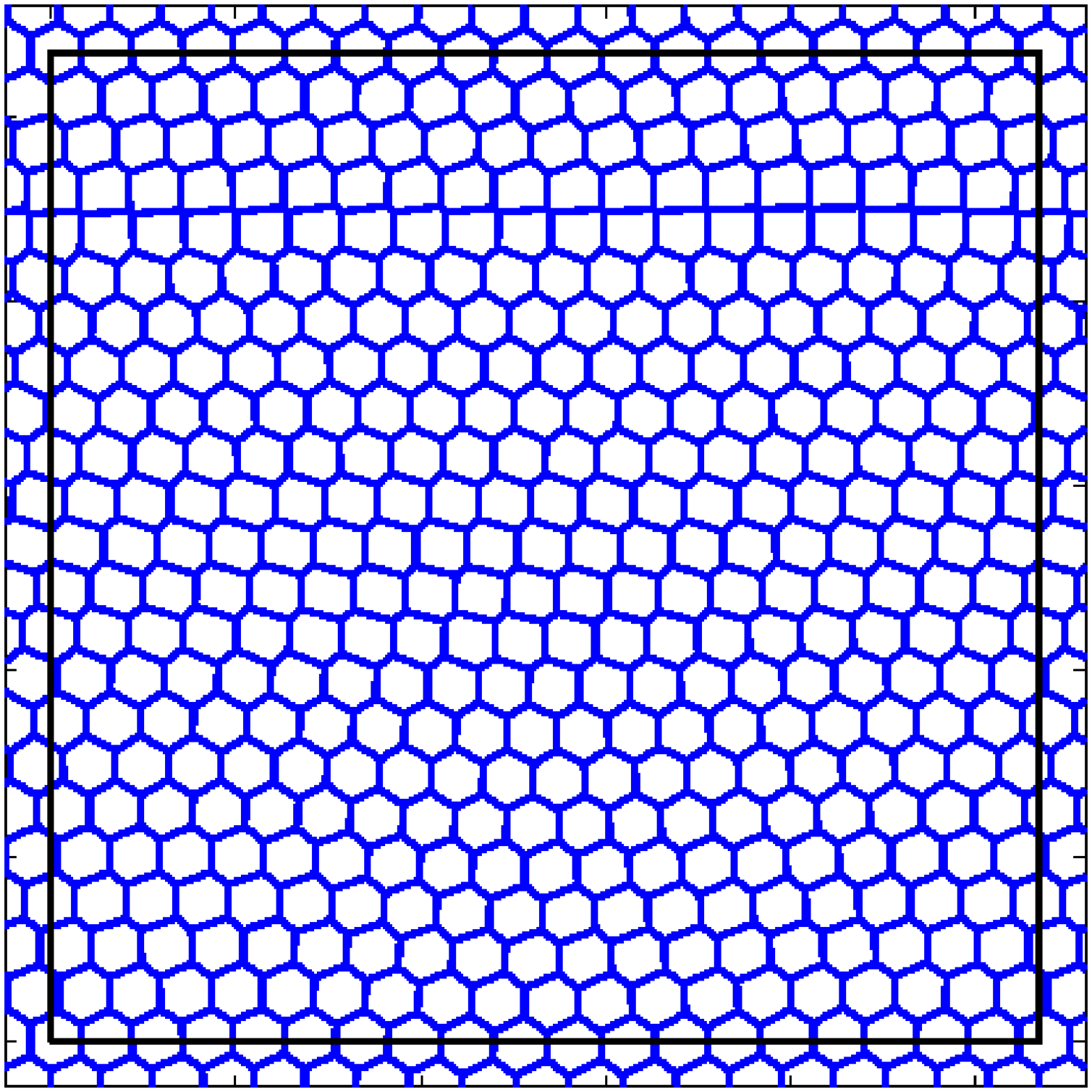}
\includegraphics[width=0.3\textwidth,clip=true]{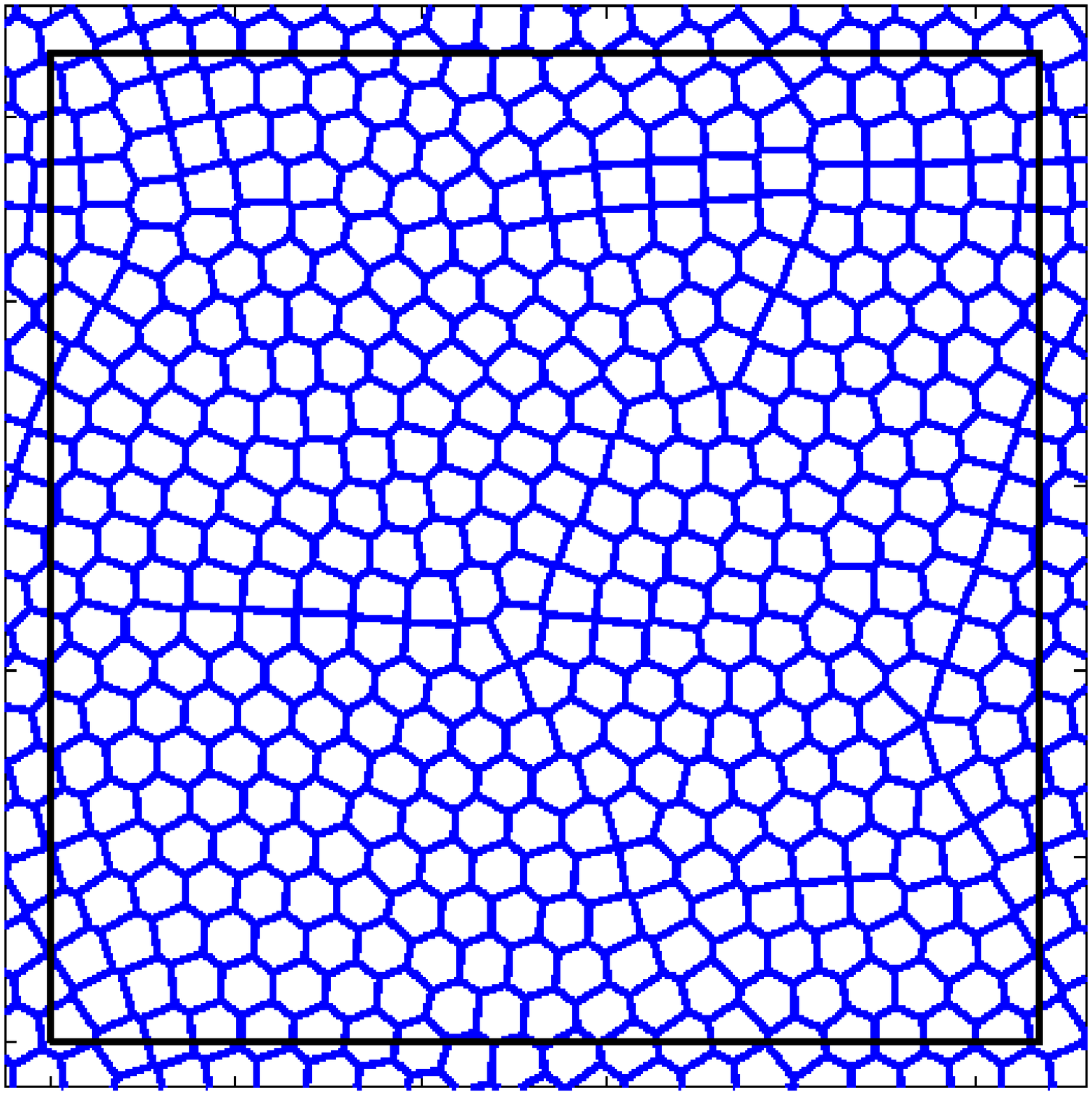}
\includegraphics[width=0.3\textwidth,clip=true]{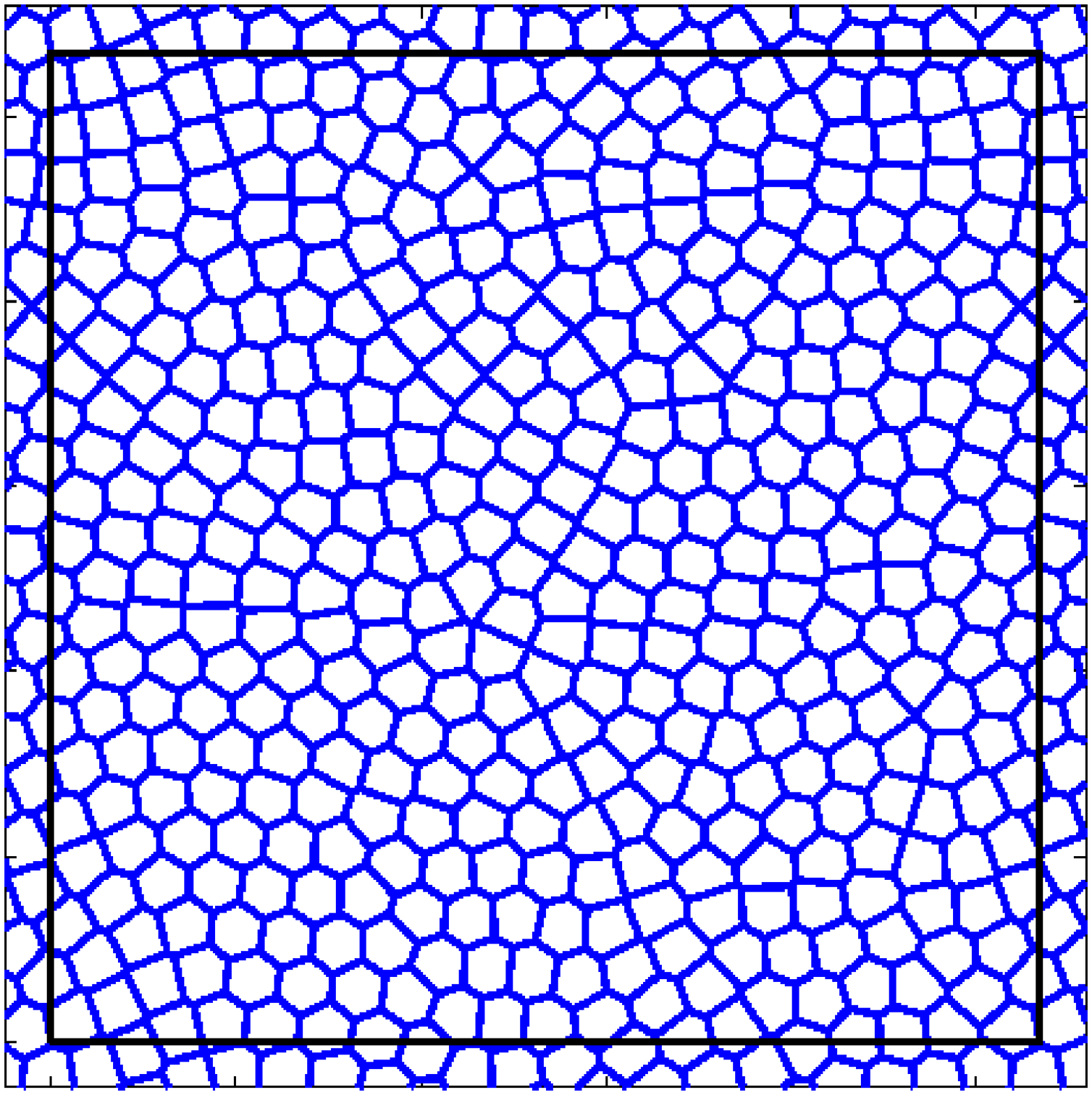}
\caption{(Color online) Configurations (top row) and associated Voronoi
tessellations (bottom row) for the minimum-energy path from ground state (left
column) through the saddle point (middle column) to the inherent structure
(right column) for $\chi=0.5383$. The transitions are from the set described in
Fig.\ \ref{fig:barriers}. }
\label{fig:path}
\end{figure}

\begin{figure}
\includegraphics[width=0.3\textwidth, clip=true]{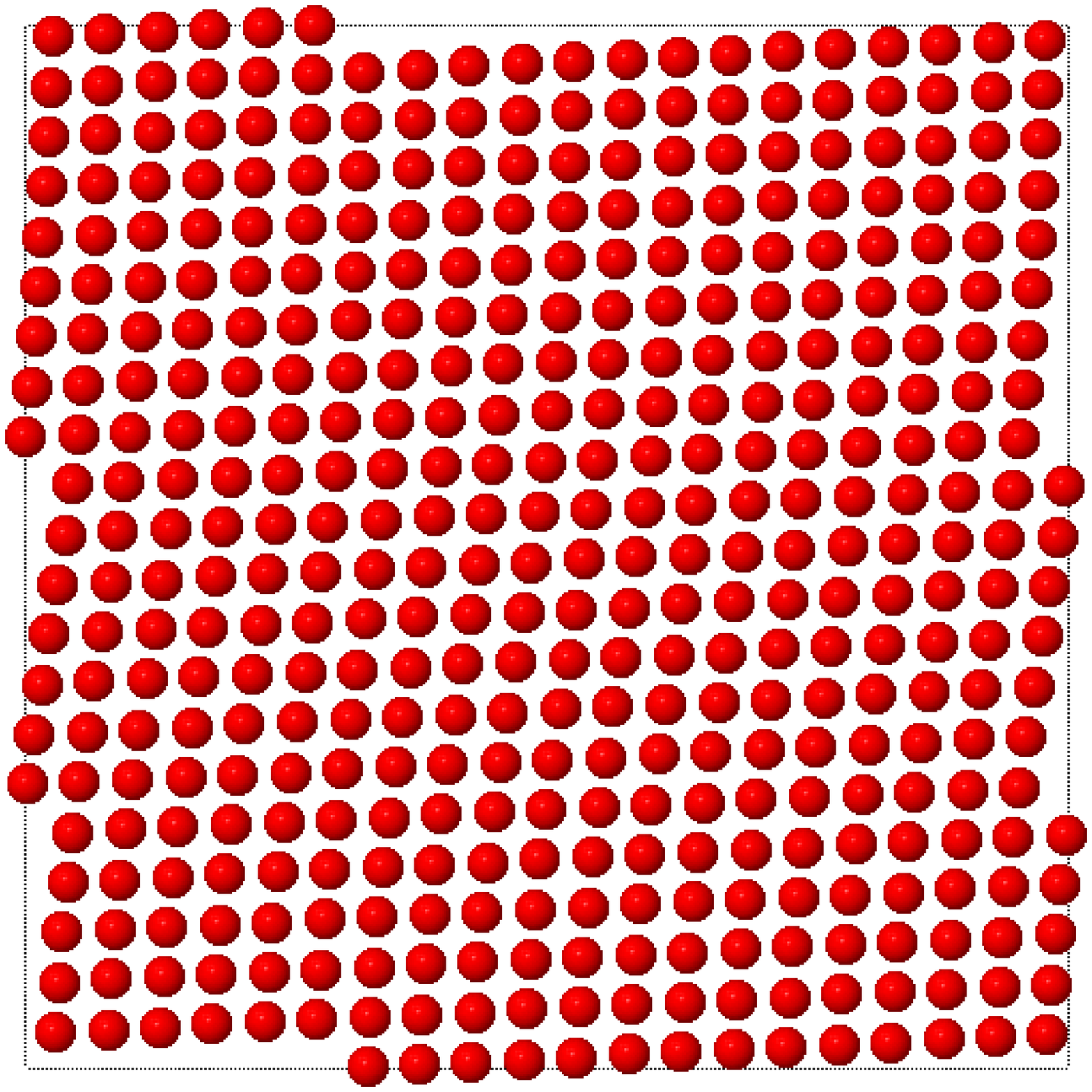}
\includegraphics[width=0.3\textwidth, clip=true]{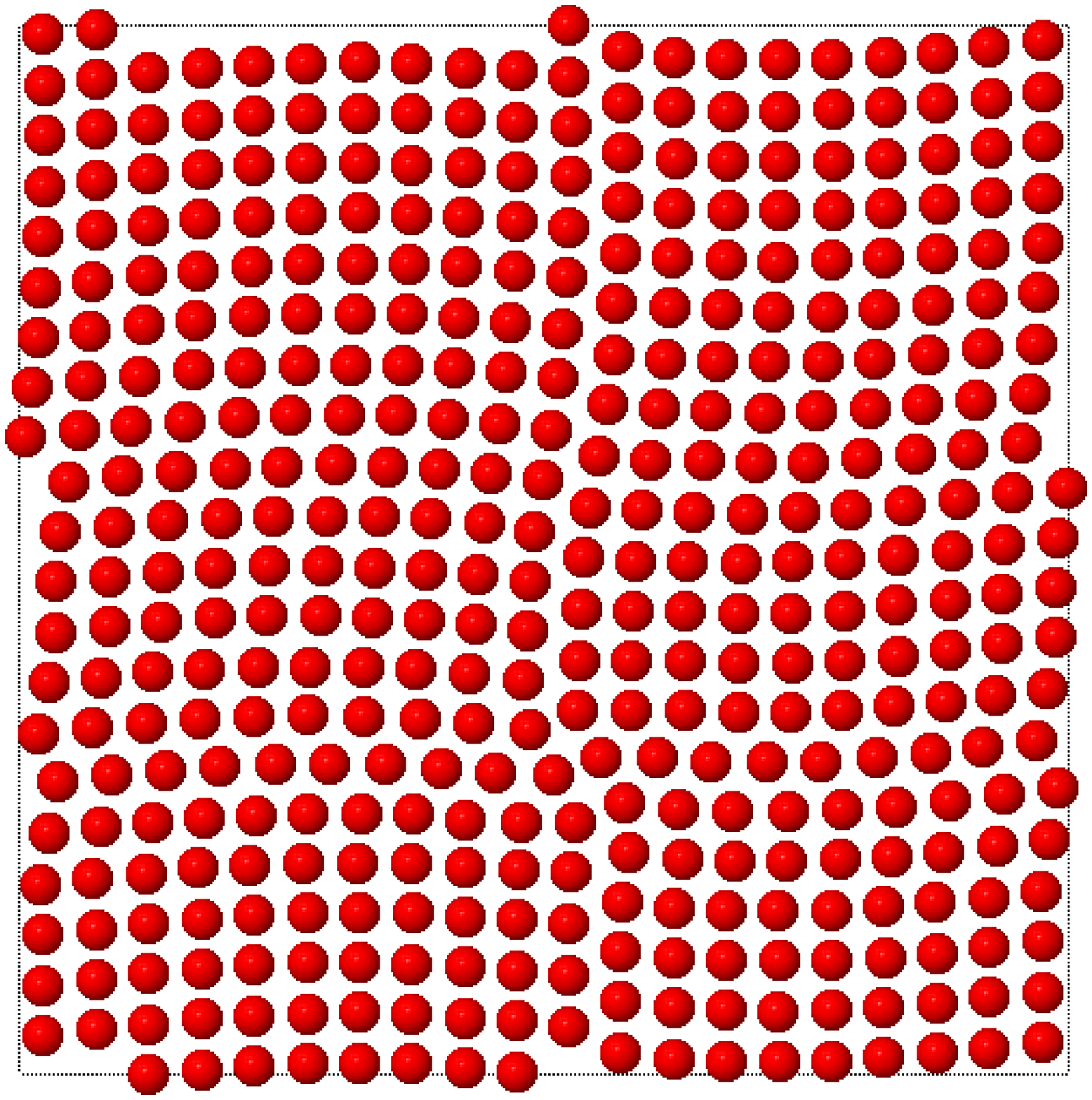}
\includegraphics[width=0.3\textwidth, clip=true]{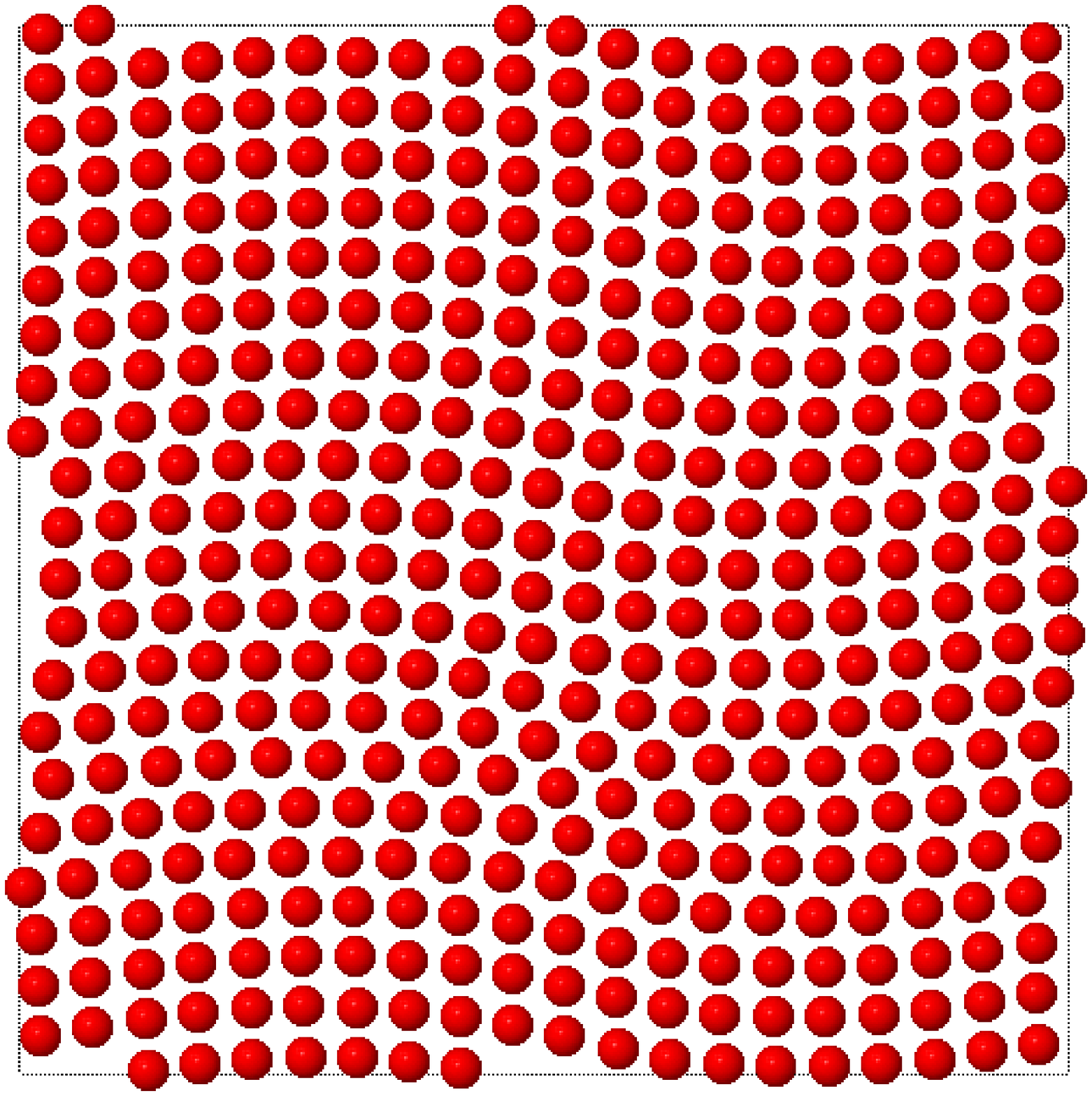} 
\includegraphics[width=0.3\textwidth, clip=true]{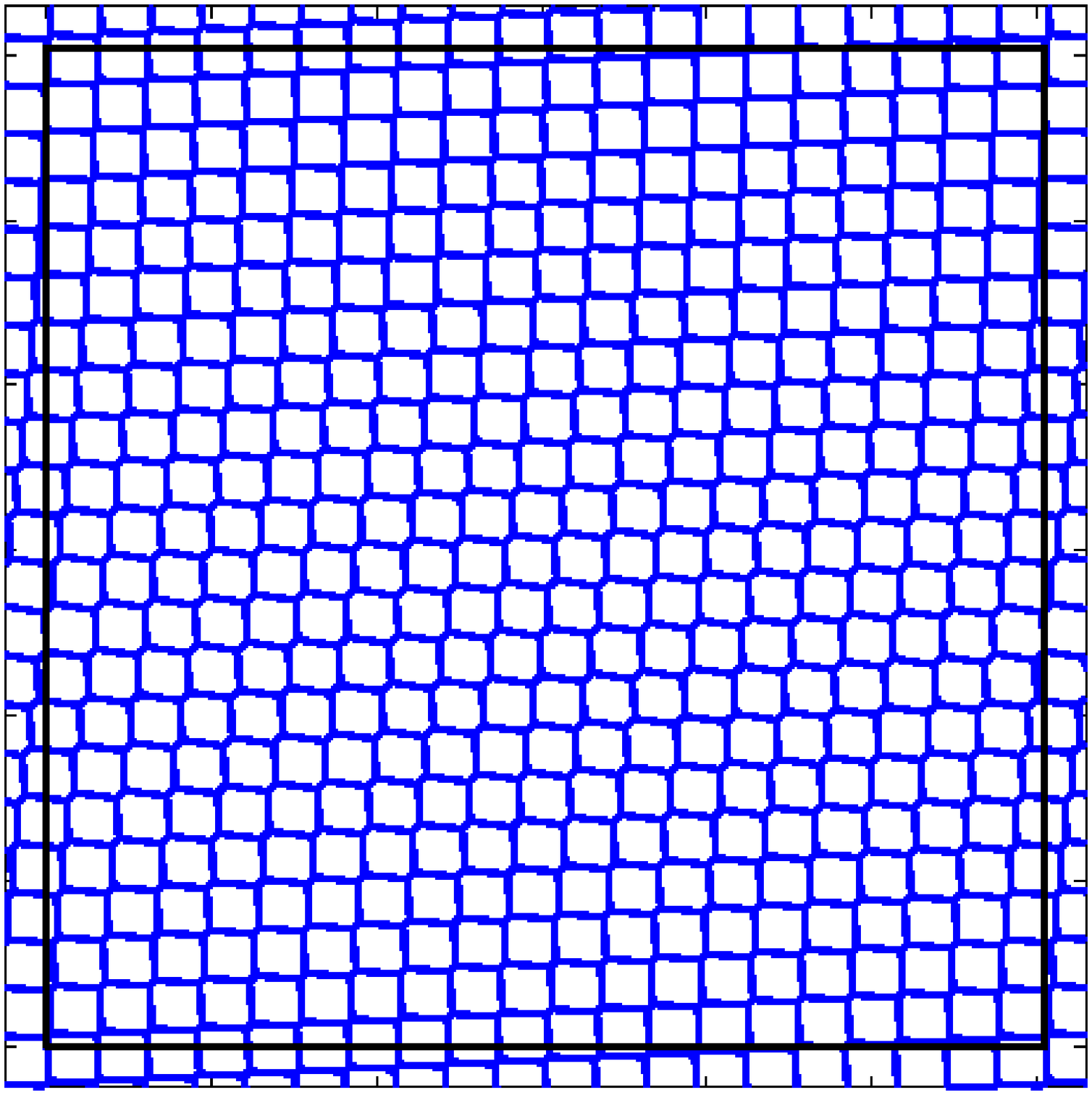}
\includegraphics[width=0.3\textwidth, clip=true]{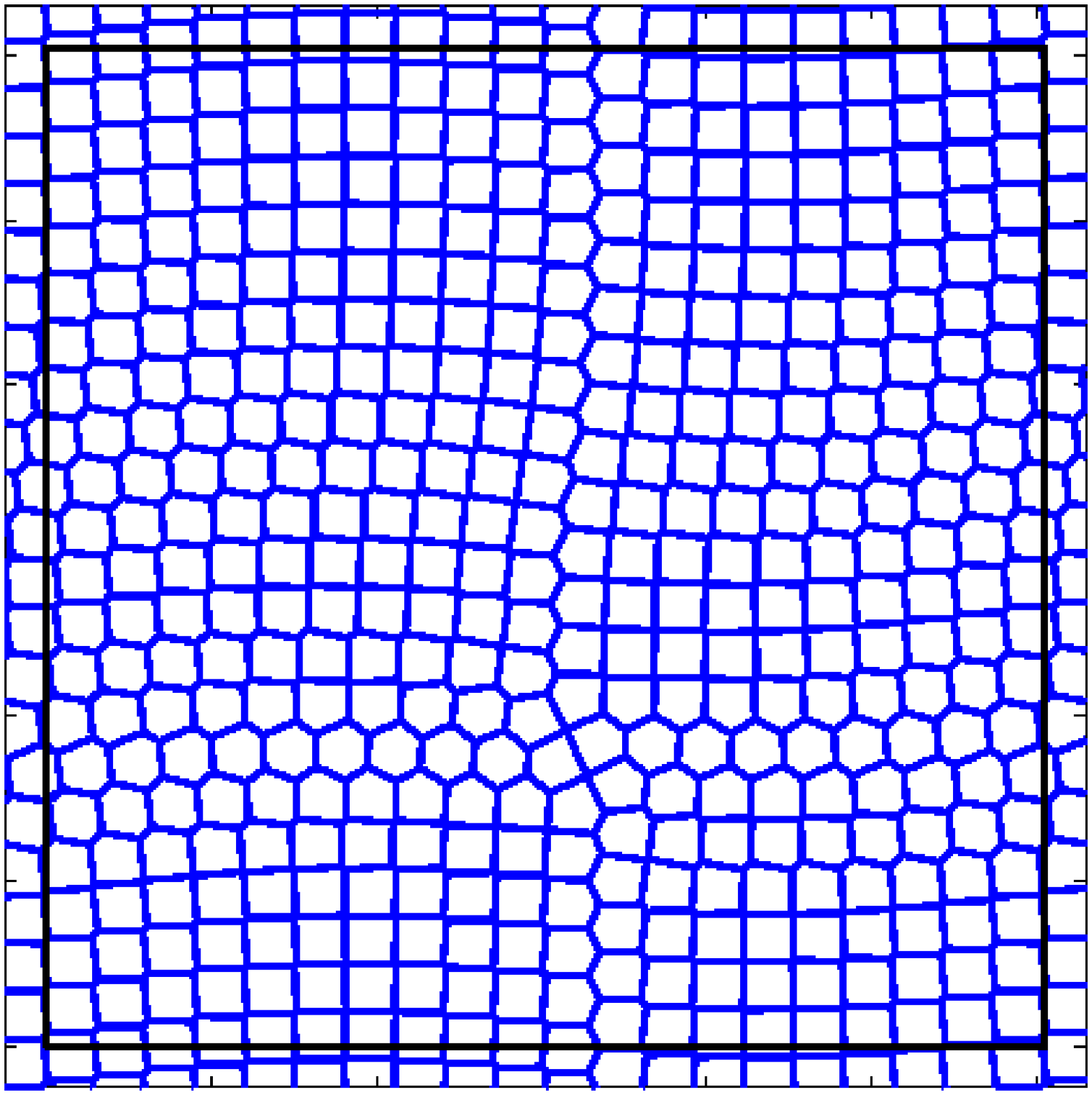}
\includegraphics[width=0.3\textwidth, clip=true]{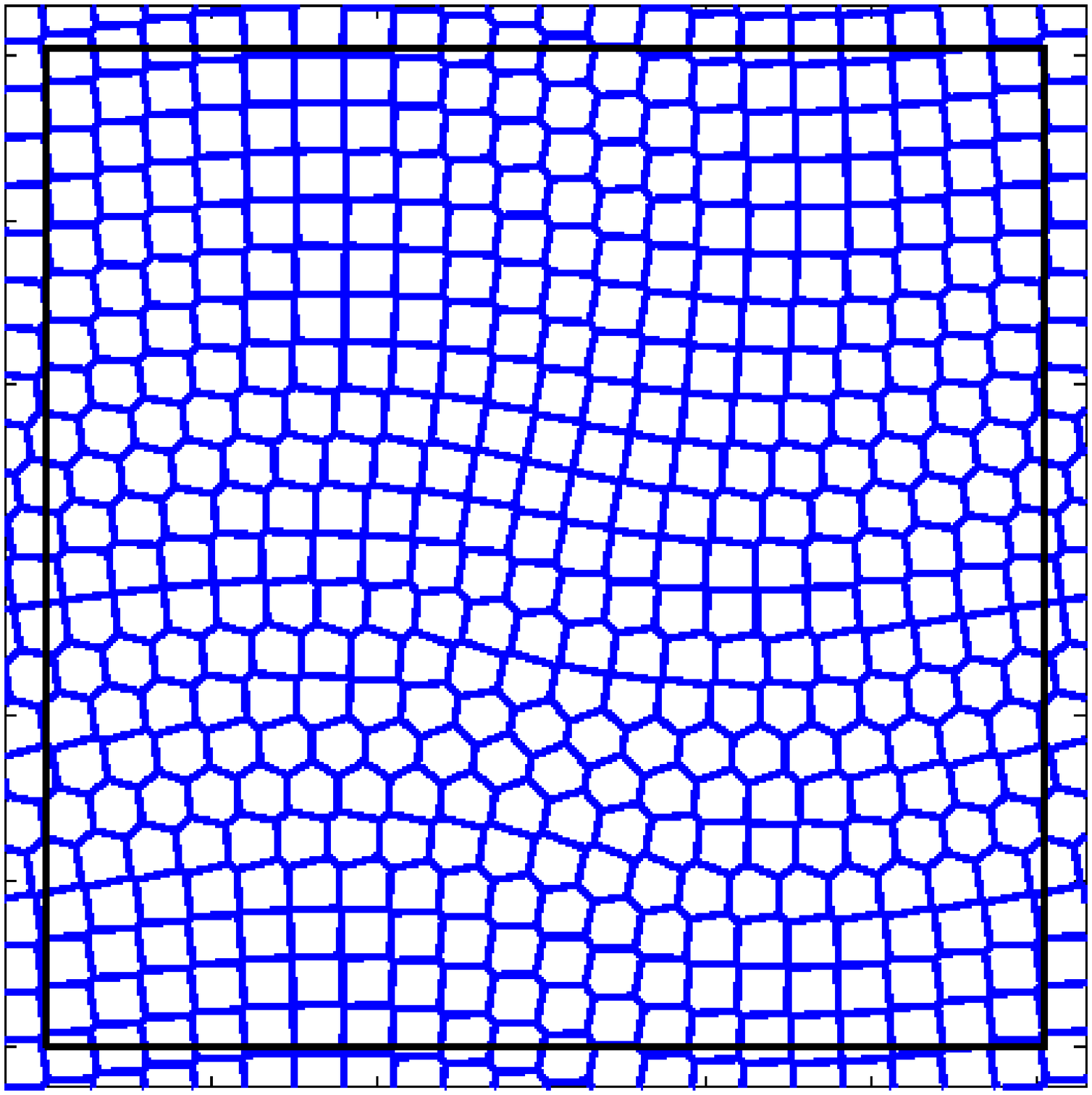}
\caption{ (Color online) Configurations (top row) and associated Voronoi
tessellations (bottom row) for the minimum-energy path from ground state (left
column) through the saddle point (middle column) to the inherent structure
(right column) for $\chi=0.7033$. The transitions are from the set described in
Fig.\ \ref{fig:barriers}. }
\label{fig:path2}
\end{figure}

In Figs.\ \ref{fig:path} and \ref{fig:path2}, we display the ground state,
saddle point, and inherent structure and the associated Voronoi diagrams for
$\chi$=0.5383 and 0.7033. These are the same paths shown in Fig.\
\ref{fig:barriers}. These systems are representative of the types of local
rearrangements found in the transition from ground states to inherent
structures.  In Fig.\ \ref{fig:path}, the ground state has wavy-crystalline
characteristics where there is alignment in one direction. The Voronoi diagram
appears to have highly ordered arrangement of polygons.  In the saddle point
images, one can see the appearance of the five-particle rings. In the inherent
structure, the defect appears to be localized, however, the defects are more
easily discerned in the Voronoi diagram.

For $\chi$=0.7033, the buildup of the grain boundary is clear in Fig.\
\ref{fig:path2}.  The ground state is wavy-crystalline and the saddle point
shows the appearance of a vertical and horizontal grain.  The inherent structure
shows the apparent mismatch between lines of particles that is responsible for
the inability to reach the ground state. The Voronoi diagrams in this case
visually display the source of the frustration in the system. While far from the
grain, the polygons in the Voronoi digram for the inherent structure are quite
regular and ordered, those near the grain boundary have much more variation in
shape.  For $\chi$=0.8995, the rearrangements found in the transition from a
perfect lattice to one with a vacancy-interstitial pair are very local (albeit collective), only
involving the few particles immediately surround the interstitial-vacancy pair.
In all cases observed, the rearrangements from ground states to inherent
structures involve only local, but collective, rearrangements (within a few particle diameters).
As highlighted here, they typically involve the formation of a grain boundary or
a five-particle ring.

\clearpage

\section{Particle Rearrangements from RSA patterns to ground states}
\label{sec:rsa}

Configurations generated by random sequential addition near the saturation
densities are known to suppress long-wavelength scattering, described as
``ultratransparent,''\cite{mattarelli2007ugc, mattarelli2010transparency} though
images of the structure factor for the RSA configuration demonstrates that it is
not hyperuniform and not absolutely stealthy.\cite{torquato2006rsa} From Ref.\
[34], as $k$ approaches zero, the structure factor approaches a value of 0.059,
corresponding to a value of the hyperuniformity parameter $C_0$ of -1.23.

\begin{figure}
\includegraphics[width=0.4\textwidth, clip=true]{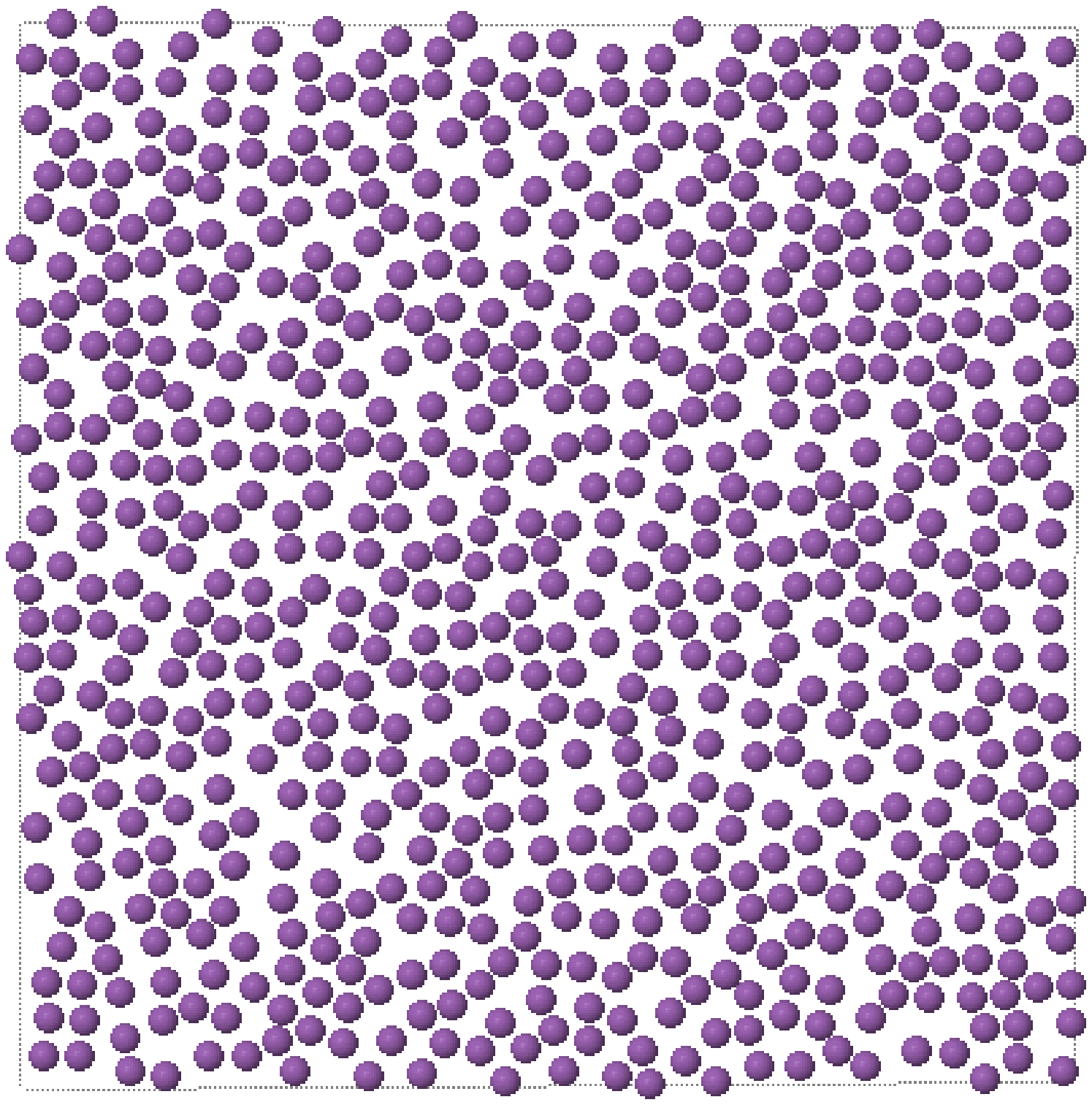}
\includegraphics[width=0.4\textwidth, clip=true]{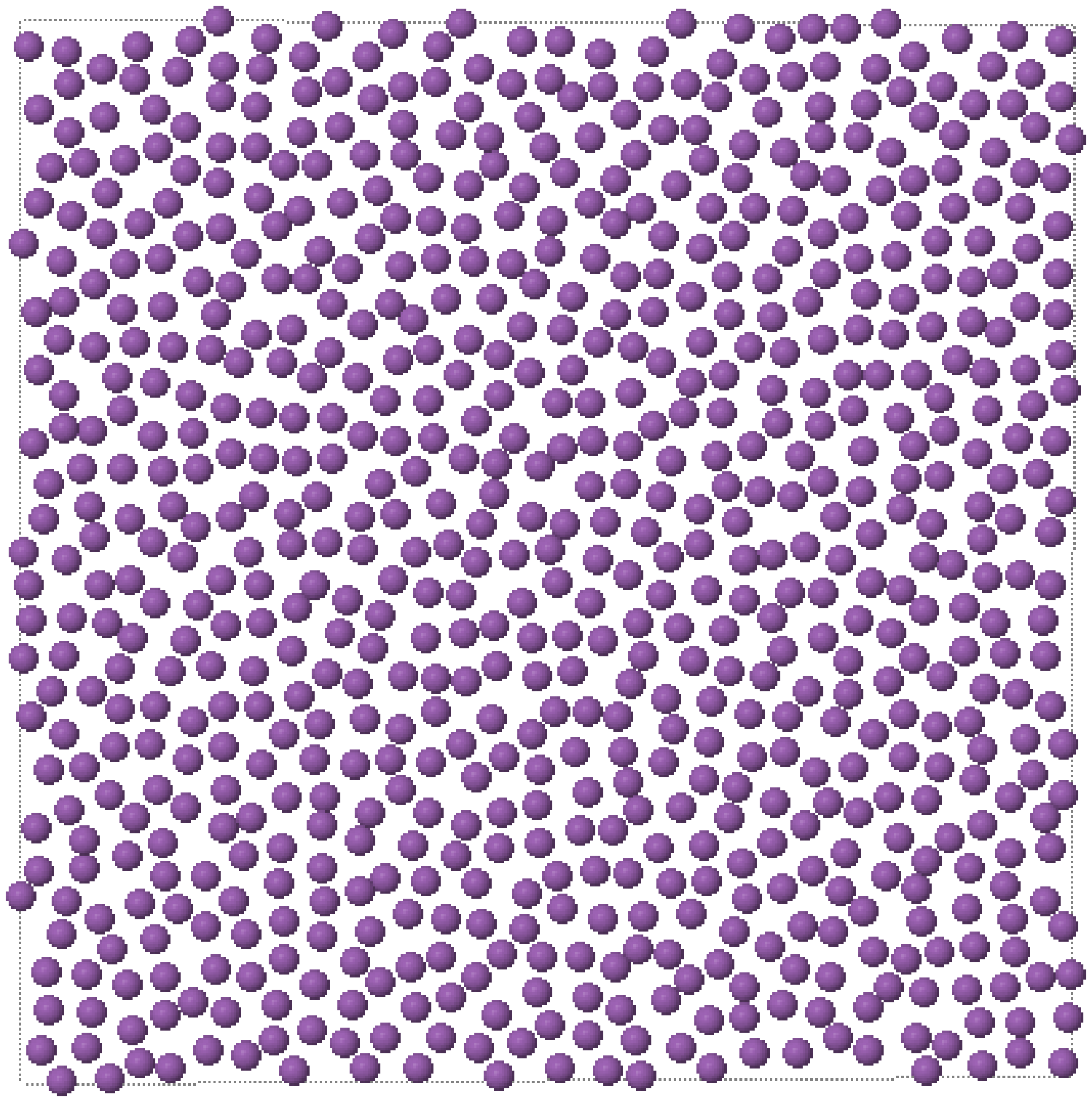} 
\includegraphics[width=0.4\textwidth, clip=true]{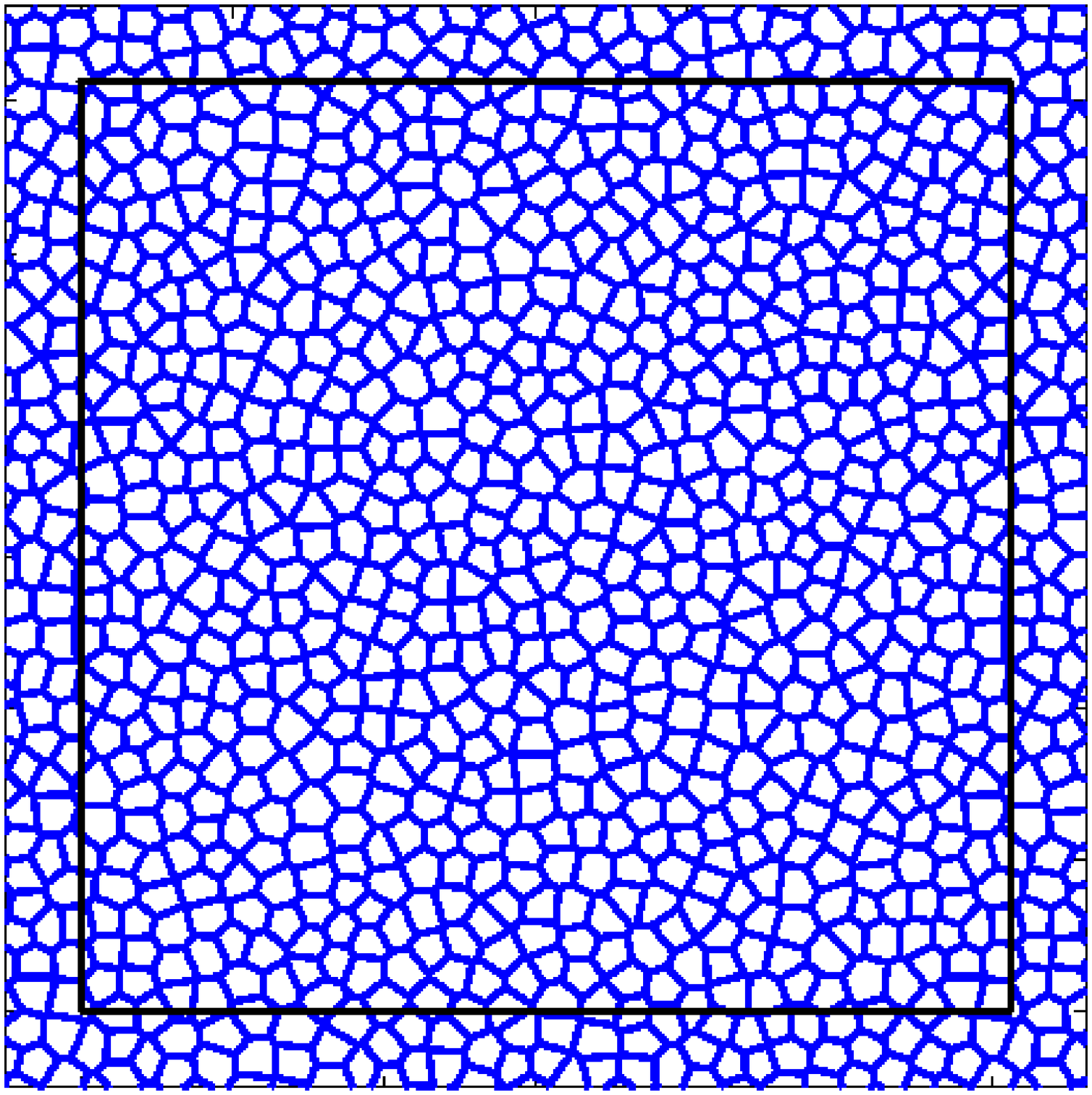}
\includegraphics[width=0.4\textwidth, clip=true]{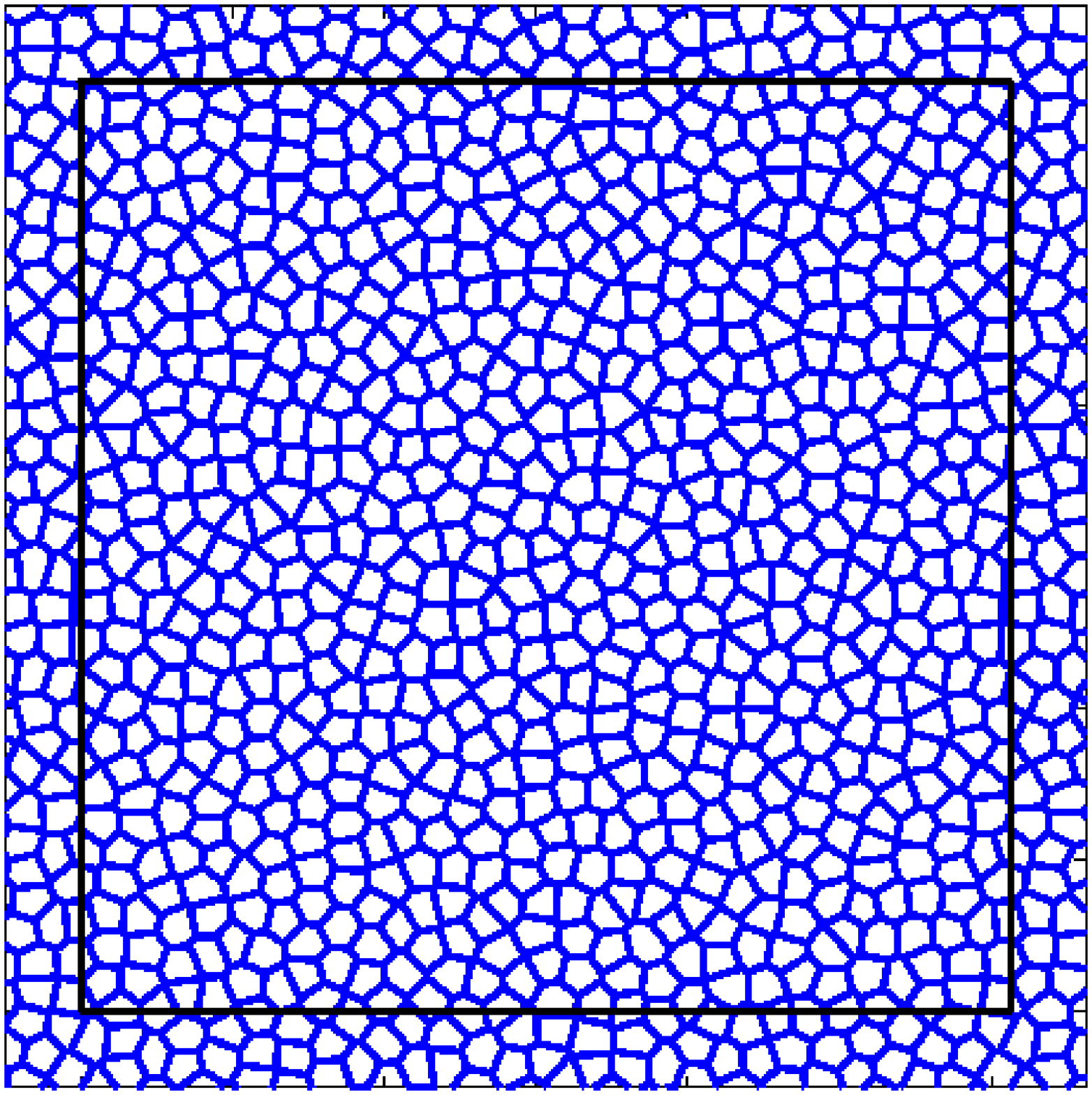} 
\caption{ (Color online)  (Top) Configurations with $\rho=0.1953$ and
$\chi=0.0991$ and (bottom) the associated Voronoi diagrams.  The left images are
the initial RSA configurations and the right images are the ground states. The
trajectory from the RSA configuration to the ground state appears to arise from
a gentle repulsion. The diameters of the particles correspond roughly to the
assigned RSA diameter. The dark lines represent the system box.}
\label{fig:rsax0991}
\end{figure}

\begin{figure}
\includegraphics[width=0.4\textwidth, clip=true]{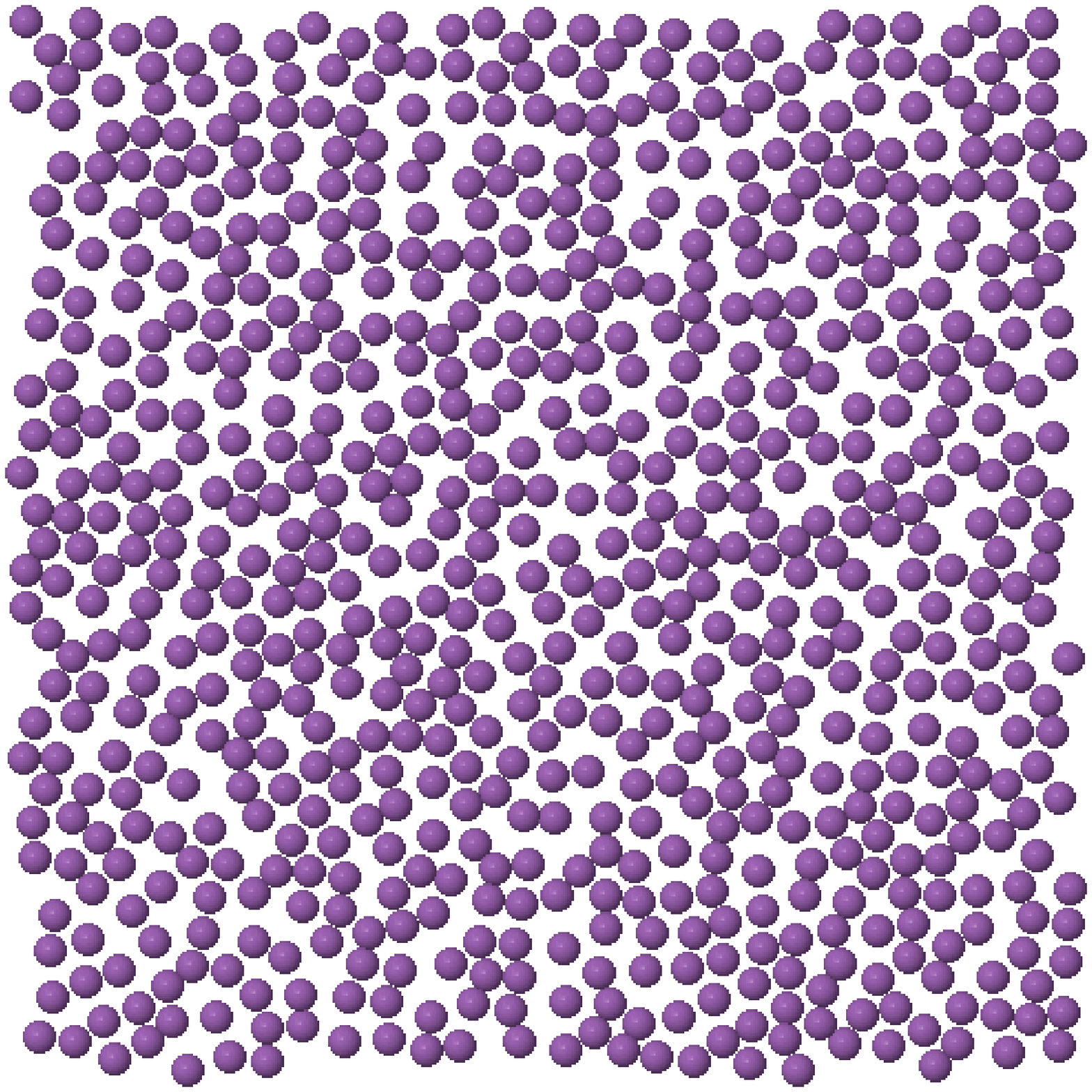}
\includegraphics[width=0.4\textwidth, clip=true]{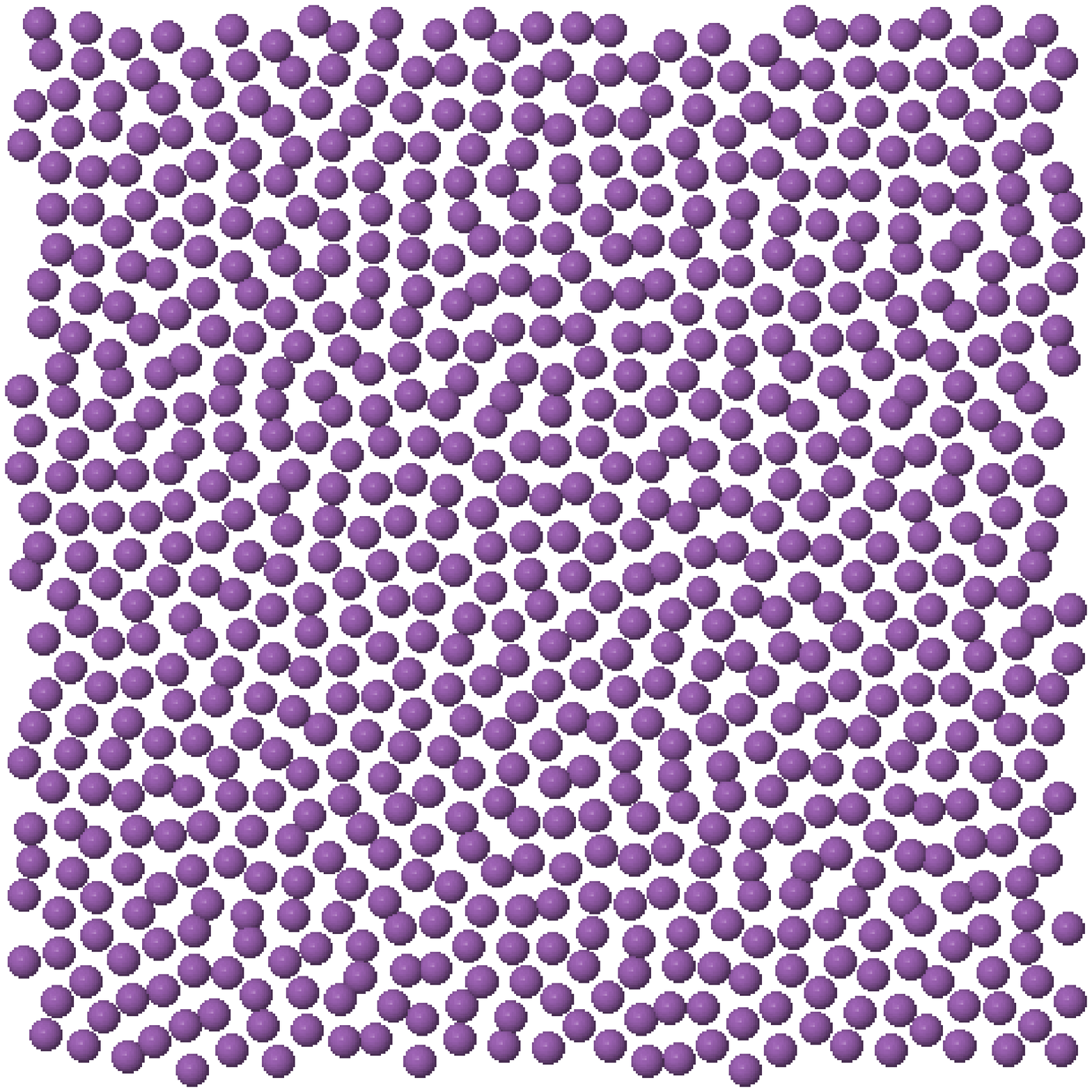}
\includegraphics[width=0.4\textwidth, clip=true]{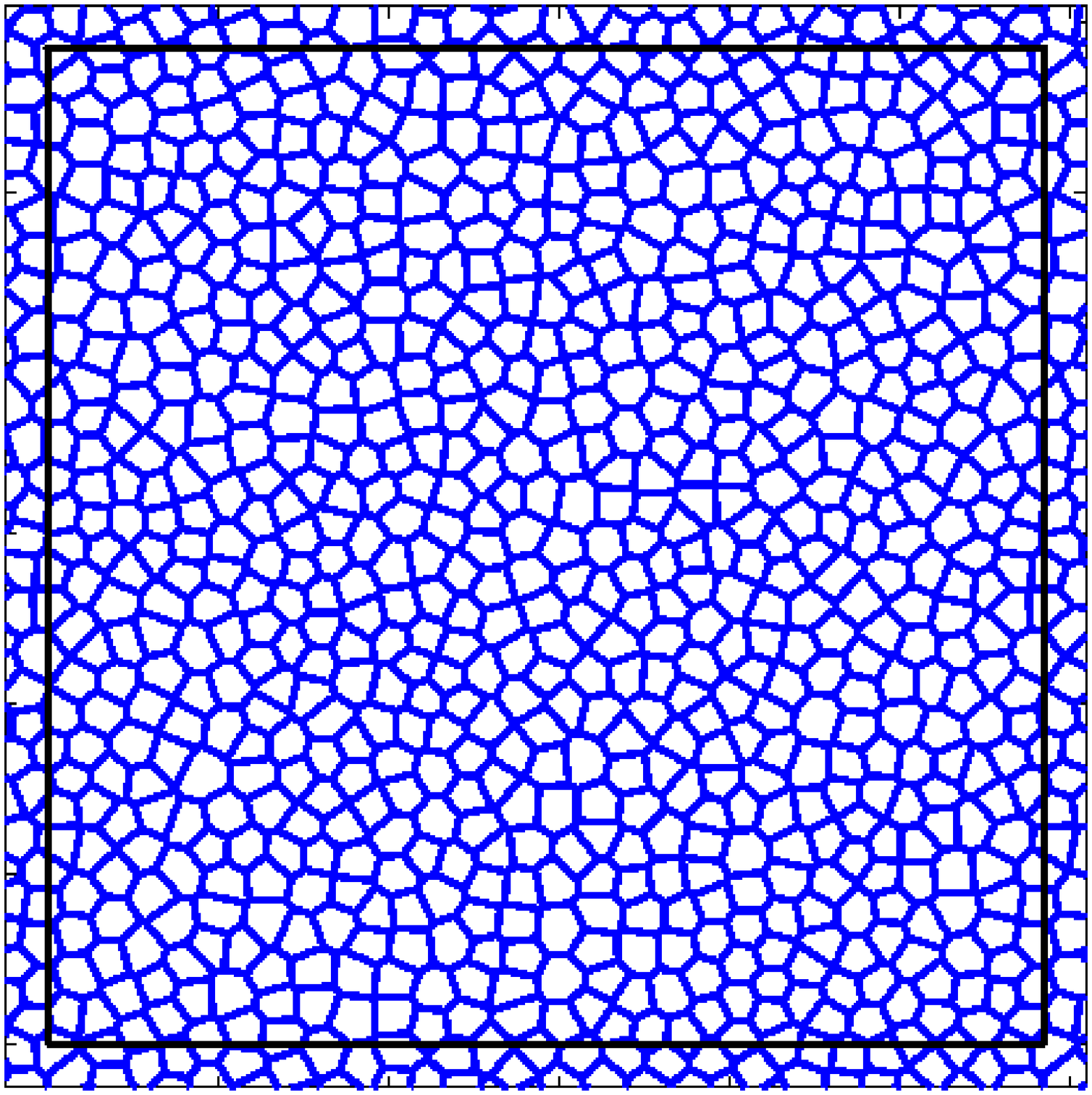}
\includegraphics[width=0.4\textwidth, clip=true]{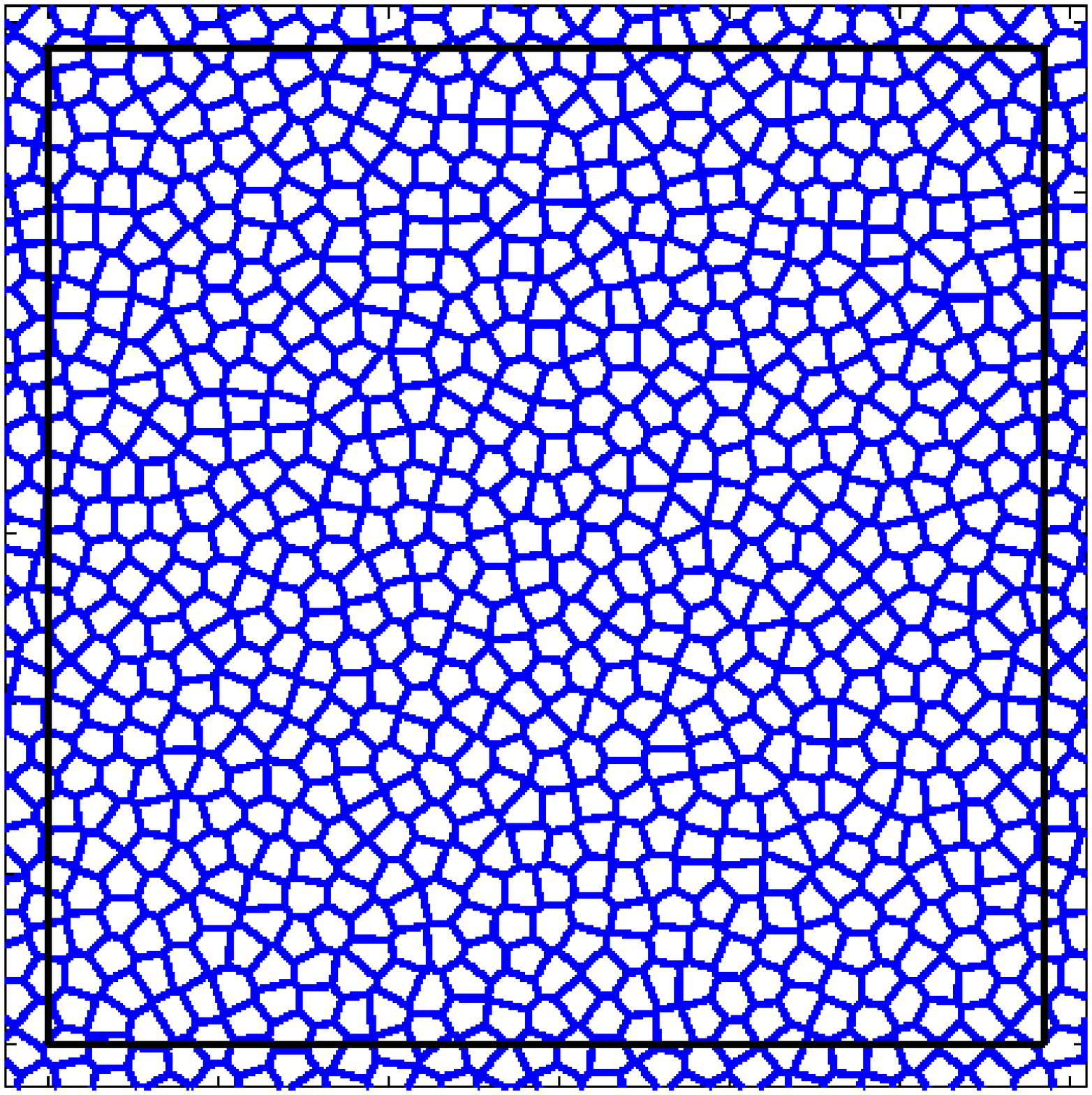} 
\caption{ (Color online)  (Top) Configurations with $\rho=0.05434$ and
$\chi=0.3657$ and (bottom) the associated Voronoi diagrams.  The left images are
the initial RSA configurations and the right images are the ground states. The
trajectory from the RSA configuration to the ground state appears to arise from
a gentle repulsion. The diameters of the particles correspond roughly to the
assigned RSA diameter. The dark lines represent the system box.}
\label{fig:rsax3657}
\end{figure}

In Figs.\ \ref{fig:rsax0991} and \ref{fig:rsax3657}, we display the initial RSA
configurations and the final ground-state configurations as obtained from the
collective coordinate approach, for $\chi$ values of 0.0991 and 0.3657.  Each
figure also displays the corresponding Voronoi tessellation derived from the
point patterns.  In both cases when observing the dynamics of the minimization
algorithm, one observes a general spreading of particles due to the repulsion
for small $r$.  These collective rearrangements are very local and there are no large
global rearrangements involved. While every particle is displaced collectively from their
initial location, each particle is displaced by a small fraction of the
mean-nearest neighbor distance, as measured by the proximity metric per
particle, $p/N$.

In Fig.\ \ref{fig:rsax0991}, one can observe similar arrangements of particles
in both the RSA configuration and the ground-state configuration. These images
would be difficult to distinguish visually.  Several structural features remain
in tact aside from the general spreading of particles.  In the Voronoi
tessellations of the RSA configuration, there are areas where the local density
appears higher than in other areas. However, in the tessellation of the ground
state, these areas of higher local density have been relaxed away. The density
appears much more uniform in the ground state than the RSA configuration. 
However, the visual difference between the two images are subtle.  In Fig.\
\ref{fig:rsax3657}, the spreading remains local, although it is clear that the
repulsion is stronger.  The particles, on average, travel a larger distance and
fewer structural features of the RSA pattern are maintained. The Voronoi
tessellation of the ground state appears to have more uniform cell sizes than
the RSA configuration, though the differences are subtle.

\clearpage

We have quantified the differences between the RSA structures and the ground
states for several RSA configurations at various $\chi$ values. Table
\ref{tab:rsa} displays the parameters of these systems as well as the
differences in potential energy $\phi$ between the RSA system and ground state,
the stealthiness metric for RSA configurations $\eta_{rsa}$, and the
configurational proximity metric $p$.  The stealthiness metric $\eta$ for ground
states vanishes identically to zero. 

There are some variations in energies and metrics that are attributed to finite
system size. Structural features of the RSA patterns can vary across samples,
and these variations show up in Table \ref{tab:rsa} particularly for small
$\chi$.  In general, the difference between the potential energy of the RSA
system and the ground state increases uniformly with $\chi$, but is on the order
of $10^{-4}$.  The stealthiness metric also increases uniformly with $\chi$ as
is expected, since decreasing the density of RSA configurations will rescale the
structure factor to be less stealthy.  The configurational proximity metric $p$
increases uniformly with $\chi$, ranging on a per particle basis of 0.00621 to
0.01478 for these $\chi$.  These metrics show that particles in the RSA
configuration need to be collectively displaced locally by a small fraction of the
mean-nearest-neighbor distance in order to produce stealthy, hyperuniform
systems. 

\begin{table}
\caption{\label{tab:rsa} Comparison of RSA configurations to their associated ground states. }
\begin{ruledtabular}
\begin{tabular}{lll|ccccc}
$\rho$ & $\chi$ & $N$ & $D$   & $\phi_{rsa}-\phi_{gs}$ & $\eta_{rsa}$  & $p$   & $p/N$ \\\hline
0.19973 & 0.0974 & 749 & 1.866 & 5.928525$\times10^{4}$                & 0.06614       & 4.653 & 0.00621 \\
0.19653 & 0.0991 & 737 & 1.883 & 5.462281$\times10^{4}$                & 0.05879       & 5.258 & 0.00713 \\ 
0.14660 & 0.1364 & 733 & 2.155 & 5.962653$\times10^{4}$                & 0.07407       & 4.915 & 0.00671 \\
0.09893 & 0.1995 & 742 & 2.639 & 6.560097$\times10^{4}$                & 0.07569       & 6.354 & 0.00856 \\
0.06851 & 0.2902 & 734 & 3.154 & 7.001039$\times10^{4}$                & 0.08924       & 6.535 & 0.00890 \\
0.05434 & 0.3657 & 741 & 3.559 & 7.501234$\times10^{4}$                & 0.09812       & 6.943 & 0.00937 \\
0.04678 & 0.4227 & 731 & 3.809 & 8.202199$\times10^{4}$                & 0.11910       & 8.342 & 0.01141 \\
0.04161 & 0.4818 & 743 & 4.072 & 8.081648$\times10^{4}$                & 0.14692       & 10.99 & 0.01478 \\
\end{tabular}
\end{ruledtabular}
\end{table}

\section{Discussion}
\label{sec:disc}

In this paper, we have shown that the energy landscape associated with the
$k$-space overlap potential is relatively flat and devoid of deep energy wells.
This is to be contrasted with Lennard-Jones systems, which possess rugged energy
landscapes. The sampling of inherent structures from the energy landscape is
independent of the temperature from which the system was sampled and the rate at
which the system was cooled. Five-particle rings, which are related to the local
interactions between particles, disrupt the ability of a system to become a
stealthy ground state, while grain boundaries, which arise for $\chi$ values in
the wavy-crystalline regime, are attributed to the longer-range interactions in
a system. While the hyperuniformity parameter $C_0$ and the stealthiness
parameter $\eta$ are positively correlated with each other, the bond-order
parameter $\Psi_6$ can display a broad range of characteristics for various
$\chi$ values. We have used the nudged-elastic-band algorithm to show that local
particle rearrangements can allow systems to be perturbed over a relatively
small energy barrier from an inherent structure to a ground state. These
rearrangements correspond to a low configurational proximity metric $p$. Lastly,
we have shown that highly local collective particle displacements, as quantified
via the configurational proximity metric $p$, are sufficient to convert a RSA 
configuration into a stealthy, hyperuniform ground state. 

While these studies represent another contribution to understanding the nature
of these unusual long-ranged potentials, there are some natural extensions to
this work that can further illuminate the relationship between particle
interactions, stealthiness, and hyperuniformity. We have used simple metrics
$C_0$ and $\eta$ as measures of stealthiness and hyperuniformity, there are a
multitude of other metrics that can be constructed.  Measuring hyperuniformity
for a single finite system is challenging due to the fluctuations and noise that
can arise in the small-$k$ behavior of $S(k)$. Ideally, one would prefer to have
a large system where the number variance in an observation window can be
measured accurately. However, we are limited in system size because of numerical
methods. While our measurement of stealthiness is directly related to the
structure factor, we assumed equal weights for all the $S(k)$, while one could
assign weights differently to favor small-$k$ scattering or near-$K$ scattering.
Currently, there are no obvious metrics that stand out as superior to those used
here. However, the development of new, improved metrics remains a potential
direction of future research.

While our paths connect ground states to inherent structures, a more generic
algorithm taking inherent structures to ground states would be valuable. For
instance, our methods relied on knowledge of the ground state and inherent
structure {\it a priori}. However, a more useful method would be to find the
nearest ground-state structure given only an inherent structure. This algorithm
would be far more powerful than the current method.  One could further quantify
the possible rearrangements in the inherent structures in order to help produce
such an algorithm that relies on trial displacements that search for small
energy barriers nearby. This could lead to a statistical analysis of the types
of rearrangements made in a system to determine which are more likely to lead to
a ground state. In addition, a more in-depth study of the rearrangements from
RSA configurations to hyperuniform and stealthy configurations would be
meaningful.  Perhaps one can distinguish the relative contributions of the
features of the interaction potential that contribute to the hyperuniformity and
stealthiness of a system.

The precise role of the pair interactions in the formation of disordered ground
states and stealthiness is unknown. Here, we have determined that five-particle
rings are a consequence of the first two minima in the overlap potential $v(r)$.
The ground-state behavior holds for a class of potentials\cite{batten2008cdg}
and we strongly suspect that similar behavior in the inherent structure analysis
may generally hold as well.  A systematic study involving the determination of
ground states for the truncated overlap potential in Eq.\ (\ref{eq:overlapvr})
might help understand how many local minima in $v(r)$ are necessary for
disordered ground states to arise. The potential could be truncated at the
first, second, third, {\it etc}.\ minimum and numerical procedures could be used
to find ground-state structures. This would encounter issues of rigor in that
the structures found by numerical methods may not touch the lower bound on
$\phi$.  The trivial lower bound for $\phi$ is zero (due to the nonnegativity of
the overlap potential), but for the densities associated with the disordered and
wavy-crystalline regimes, the improved lower bound, that being $S(k)$ must
vanish for all $k<K$, is not applicable to the truncated $k$-space overlap
potential.

In addition, we observed that the shape of $V(k)$, which assigns weights to
$S(k)$ in the potential-energy function, apparently influences the shape of the
function $S(k)$ for the inherent structures for $\chi<0.58$.  Above this $\chi$
value, there was no obvious connection between $V(k)$ and $S(k)$ for the
inherent structures. This raises the question as to whether the shape of $V(k)$
actually influences the shape of the function $S(k)$ for inherent structures. A
simple study varying the shape of $V(k)$, but maintaining the compact support at
$K$ and the positivity of $V(k)$ would help answer this question.

Lastly, while this collective coordinate setup was used understand and design
new materials in two-dimensions, such as photonic bandgap
materials,\cite{florescu2009designer} the extension to three dimensions should
be particularly fruitful. While we have some understanding of stealthy ground
states in three dimensions,\cite{uche2006ccc, batten2008cdg} the relations
between inherent structures and ground states may be different in three
dimensions than two dimensions. Connecting these structural rearrangements and
physical properties can help to contribute to the next class of
three-dimensional photonic bandgap materials.

\section*{Acknowledgments}
This research was supported by the U.S. Department of Energy, Office of
Basic Energy Sciences, Division of Materials Sciences and Engineering
under Award DE-FG02-04-ER46108.


\newpage

\begin{thebibliography}{44}
\expandafter\ifx\csname natexlab\endcsname\relax\def\natexlab#1{#1}\fi
\expandafter\ifx\csname bibnamefont\endcsname\relax
  \def\bibnamefont#1{#1}\fi
\expandafter\ifx\csname bibfnamefont\endcsname\relax
  \def\bibfnamefont#1{#1}\fi
\expandafter\ifx\csname citenamefont\endcsname\relax
  \def\citenamefont#1{#1}\fi
\expandafter\ifx\csname url\endcsname\relax
  \def\url#1{\texttt{#1}}\fi
\expandafter\ifx\csname urlprefix\endcsname\relax\def\urlprefix{URL }\fi
\providecommand{\bibinfo}[2]{#2}
\providecommand{\eprint}[2][]{\url{#2}}

\bibitem[{\citenamefont{Batten et~al.}(2009{\natexlab{a}})\citenamefont{Batten,
  Stillinger, and Torquato}}]{batten2009novel}
\bibinfo{author}{\bibfnamefont{R.~D.} \bibnamefont{Batten}},
  \bibinfo{author}{\bibfnamefont{F.~H.} \bibnamefont{Stillinger}},
  \bibnamefont{and} \bibinfo{author}{\bibfnamefont{S.}~\bibnamefont{Torquato}},
  \bibinfo{journal}{Phys. Rev. Lett} \textbf{\bibinfo{volume}{103}},
  \bibinfo{pages}{50602} (\bibinfo{year}{2009}{\natexlab{a}}).

\bibitem[{\citenamefont{Batten et~al.}(2009{\natexlab{b}})\citenamefont{Batten,
  Stillinger, and Torquato}}]{batten2009interactions}
\bibinfo{author}{\bibfnamefont{R.~D.} \bibnamefont{Batten}},
  \bibinfo{author}{\bibfnamefont{F.~H.} \bibnamefont{Stillinger}},
  \bibnamefont{and} \bibinfo{author}{\bibfnamefont{S.}~\bibnamefont{Torquato}},
  \bibinfo{journal}{Phys. Rev. E} \textbf{\bibinfo{volume}{80}},
  \bibinfo{pages}{31105} (\bibinfo{year}{2009}{\natexlab{b}}).

\bibitem[{\citenamefont{Batten et~al.}(2008)\citenamefont{Batten, Stillinger,
  and Torquato}}]{batten2008cdg}
\bibinfo{author}{\bibfnamefont{R.~D.} \bibnamefont{Batten}},
  \bibinfo{author}{\bibfnamefont{F.~H.} \bibnamefont{Stillinger}},
  \bibnamefont{and} \bibinfo{author}{\bibfnamefont{S.}~\bibnamefont{Torquato}},
  \bibinfo{journal}{J. Appl. Phys.} \textbf{\bibinfo{volume}{104}},
  \bibinfo{eid}{033504} (\bibinfo{year}{2008}).

\bibitem[{\citenamefont{Uche et~al.}(2006)\citenamefont{Uche, Torquato, and
  Stillinger}}]{uche2006ccc}
\bibinfo{author}{\bibfnamefont{O.~U.} \bibnamefont{Uche}},
  \bibinfo{author}{\bibfnamefont{S.}~\bibnamefont{Torquato}}, \bibnamefont{and}
  \bibinfo{author}{\bibfnamefont{F.~H.} \bibnamefont{Stillinger}},
  \bibinfo{journal}{Phys. Rev. E} \textbf{\bibinfo{volume}{74}},
  \bibinfo{pages}{31104} (\bibinfo{year}{2006}).

\bibitem[{\citenamefont{Torquato and Stillinger}(2008)}]{torquato2008ndr}
\bibinfo{author}{\bibfnamefont{S.}~\bibnamefont{Torquato}} \bibnamefont{and}
  \bibinfo{author}{\bibfnamefont{F.~H.} \bibnamefont{Stillinger}},
  \bibinfo{journal}{Phys. Rev. Lett.} \textbf{\bibinfo{volume}{100}},
  \bibinfo{pages}{020602} (\bibinfo{year}{2008}).

\bibitem[{\citenamefont{Torquato}(2009)}]{torquato2009iot}
\bibinfo{author}{\bibfnamefont{S.}~\bibnamefont{Torquato}},
  \bibinfo{journal}{Soft Matter} \textbf{\bibinfo{volume}{5}},
  \bibinfo{pages}{1157} (\bibinfo{year}{2009}).

\bibitem[{\citenamefont{Fan et~al.}(1991)\citenamefont{Fan, Percus, Stillinger,
  and Stillinger}}]{fan1991ccd}
\bibinfo{author}{\bibfnamefont{Y.}~\bibnamefont{Fan}},
  \bibinfo{author}{\bibfnamefont{J.~K.} \bibnamefont{Percus}},
  \bibinfo{author}{\bibfnamefont{D.~K.} \bibnamefont{Stillinger}},
  \bibnamefont{and} \bibinfo{author}{\bibfnamefont{F.~H.}
  \bibnamefont{Stillinger}}, \bibinfo{journal}{Phys. Rev. A}
  \textbf{\bibinfo{volume}{44}}, \bibinfo{pages}{2394} (\bibinfo{year}{1991}).

\bibitem[{\citenamefont{Uche et~al.}(2004)\citenamefont{Uche, Stillinger, and
  Torquato}}]{uche2004ccd}
\bibinfo{author}{\bibfnamefont{O.~U.} \bibnamefont{Uche}},
  \bibinfo{author}{\bibfnamefont{F.~H.} \bibnamefont{Stillinger}},
  \bibnamefont{and} \bibinfo{author}{\bibfnamefont{S.}~\bibnamefont{Torquato}},
  \bibinfo{journal}{Phys. Rev. E} \textbf{\bibinfo{volume}{70}},
  \bibinfo{pages}{46122} (\bibinfo{year}{2004}).

\bibitem[{\citenamefont{Zachary and Torquato}(2011)}]{zachary2011inpress}
\bibinfo{author}{\bibfnamefont{C.~E.} \bibnamefont{Zachary}} \bibnamefont{and}
  \bibinfo{author}{\bibfnamefont{S.}~\bibnamefont{Torquato}},
  \bibinfo{journal}{Phys. Rev. E, in press,}  (\bibinfo{year}{2011}).

\bibitem[{\citenamefont{Torquato and Stillinger}(2003)}]{torquato2003ldf}
\bibinfo{author}{\bibfnamefont{S.}~\bibnamefont{Torquato}} \bibnamefont{and}
  \bibinfo{author}{\bibfnamefont{F.~H.} \bibnamefont{Stillinger}},
  \bibinfo{journal}{Phys. Rev. E} \textbf{\bibinfo{volume}{68}},
  \bibinfo{pages}{41113} (\bibinfo{year}{2003}).

\bibitem[{\citenamefont{Zachary and Torquato}(2009)}]{zachary2011statmech}
\bibinfo{author}{\bibfnamefont{C.~E.} \bibnamefont{Zachary}} \bibnamefont{and}
  \bibinfo{author}{\bibfnamefont{S.}~\bibnamefont{Torquato}},
  \bibinfo{journal}{J. Stat. Mechanics: Theory and Exp., P12015,}
  (\bibinfo{year}{2009}).

\bibitem[{\citenamefont{Zachary et~al.}(2009)\citenamefont{Zachary, Jiao, and
  Torquato}}]{zachary2011prl}
\bibinfo{author}{\bibfnamefont{C.~E.} \bibnamefont{Zachary}},
  \bibinfo{author}{\bibfnamefont{Y.}~\bibnamefont{Jiao}}, \bibnamefont{and}
  \bibinfo{author}{\bibfnamefont{S.}~\bibnamefont{Torquato}},
  \bibinfo{journal}{Phys. Rev. Lett., in press,}  (\bibinfo{year}{2009}).

\bibitem[{\citenamefont{Florescu et~al.}(2009)\citenamefont{Florescu, Torquato,
  and Steinhardt}}]{florescu2009designer}
\bibinfo{author}{\bibfnamefont{M.}~\bibnamefont{Florescu}},
  \bibinfo{author}{\bibfnamefont{S.}~\bibnamefont{Torquato}}, \bibnamefont{and}
  \bibinfo{author}{\bibfnamefont{P.}~\bibnamefont{Steinhardt}},
  \bibinfo{journal}{Proc. Nat. Acad.  Sci.}
  \textbf{\bibinfo{volume}{106}}, \bibinfo{pages}{20658}
  (\bibinfo{year}{2009}).

\bibitem[{\citenamefont{Man et~al.}(2010)\citenamefont{Man, Florescu,
  Matsuyama, Yadak, Torquato, Steinhardt, and Chaikin}}]{man2010experimental}
\bibinfo{author}{\bibfnamefont{W.}~\bibnamefont{Man}},
  \bibinfo{author}{\bibfnamefont{M.}~\bibnamefont{Florescu}},
  \bibinfo{author}{\bibfnamefont{K.}~\bibnamefont{Matsuyama}},
  \bibinfo{author}{\bibfnamefont{P.}~\bibnamefont{Yadak}},
  \bibinfo{author}{\bibfnamefont{S.}~\bibnamefont{Torquato}},
  \bibinfo{author}{\bibfnamefont{P.~J.} \bibnamefont{Steinhardt}},
  \bibnamefont{and} \bibinfo{author}{\bibfnamefont{P.}~\bibnamefont{Chaikin}},
  in \emph{\bibinfo{booktitle}{Conference on Lasers and Electro-Optics}}
  (\bibinfo{organization}{Optical Society of America}, \bibinfo{year}{2010}).

\bibitem[{\citenamefont{Debenedetti}(1996)}]{debenedetti1996metastable}
\bibinfo{author}{\bibfnamefont{P.~G.} \bibnamefont{Debenedetti}},
  \emph{\bibinfo{title}{{Metastable liquids: concepts and principles}}}
  (\bibinfo{publisher}{Princeton Univversity}, \bibinfo{year}{1996}).

\bibitem[{\citenamefont{Goldstein}(1969)}]{goldstein1969viscous}
\bibinfo{author}{\bibfnamefont{M.}~\bibnamefont{Goldstein}},
  \bibinfo{journal}{J. Chem. Phys.} \textbf{\bibinfo{volume}{51}},
  \bibinfo{pages}{3728} (\bibinfo{year}{1969}).

\bibitem[{\citenamefont{Debenedetti and
  Stillinger}(2001)}]{debenedetti2001supercooled}
\bibinfo{author}{\bibfnamefont{P.~G.} \bibnamefont{Debenedetti}}
  \bibnamefont{and} \bibinfo{author}{\bibfnamefont{F.~H.}
  \bibnamefont{Stillinger}}, \bibinfo{journal}{Nature}
  \textbf{\bibinfo{volume}{410}}, \bibinfo{pages}{259} (\bibinfo{year}{2001}).

\bibitem[{\citenamefont{Sastry et~al.}(1998)\citenamefont{Sastry, Debenedetti,
  and Stillinger}}]{sastry1998signatures}
\bibinfo{author}{\bibfnamefont{S.}~\bibnamefont{Sastry}},
  \bibinfo{author}{\bibfnamefont{P.~G.} \bibnamefont{Debenedetti}},
  \bibnamefont{and} \bibinfo{author}{\bibfnamefont{F.~H.}
  \bibnamefont{Stillinger}}, \bibinfo{journal}{Nature}
  \textbf{\bibinfo{volume}{393}}, \bibinfo{pages}{554} (\bibinfo{year}{1998}).

\bibitem[{\citenamefont{Stillinger and
  Weber}(1983{\natexlab{a}})}]{stillinger1983dynamics}
\bibinfo{author}{\bibfnamefont{F.~H.} \bibnamefont{Stillinger}}
  \bibnamefont{and} \bibinfo{author}{\bibfnamefont{T.~A.} \bibnamefont{Weber}},
  \bibinfo{journal}{Phys. Rev. A} \textbf{\bibinfo{volume}{28}},
  \bibinfo{pages}{2408} (\bibinfo{year}{1983}{\natexlab{a}}).

\bibitem[{\citenamefont{Stillinger and Weber}(1984)}]{stillinger1984point}
\bibinfo{author}{\bibfnamefont{F.~H.} \bibnamefont{Stillinger}}
  \bibnamefont{and} \bibinfo{author}{\bibfnamefont{T.~A.} \bibnamefont{Weber}},
  \bibinfo{journal}{J. Chem. Phys.} \textbf{\bibinfo{volume}{81}},
  \bibinfo{pages}{5095} (\bibinfo{year}{1984}).

\bibitem[{\citenamefont{Heuer}(1997)}]{heuer1997properties}
\bibinfo{author}{\bibfnamefont{A.}~\bibnamefont{Heuer}},
  \bibinfo{journal}{Phys. Rev. Lett.} \textbf{\bibinfo{volume}{78}},
  \bibinfo{pages}{4051} (\bibinfo{year}{1997}).

\bibitem[{\citenamefont{Stillinger and
  Weber}(1983{\natexlab{b}})}]{stillinger1983inherent}
\bibinfo{author}{\bibfnamefont{F.~H.} \bibnamefont{Stillinger}}
  \bibnamefont{and} \bibinfo{author}{\bibfnamefont{T.~A.} \bibnamefont{Weber}},
  \bibinfo{journal}{J. Chem. Phys.} \textbf{\bibinfo{volume}{87}},
  \bibinfo{pages}{2833} (\bibinfo{year}{1983}{\natexlab{b}}).

\bibitem[{\citenamefont{Weber and
  Stillinger}(1985{\natexlab{a}})}]{weber1985local}
\bibinfo{author}{\bibfnamefont{T.~A.} \bibnamefont{Weber}} \bibnamefont{and}
  \bibinfo{author}{\bibfnamefont{F.~H.} \bibnamefont{Stillinger}},
  \bibinfo{journal}{Phys. Rev. B} \textbf{\bibinfo{volume}{31}},
  \bibinfo{pages}{1954} (\bibinfo{year}{1985}{\natexlab{a}}).

\bibitem[{\citenamefont{Weber and
  Stillinger}(1985{\natexlab{b}})}]{weber1985interactions}
\bibinfo{author}{\bibfnamefont{T.}~\bibnamefont{Weber}} \bibnamefont{and}
  \bibinfo{author}{\bibfnamefont{F.}~\bibnamefont{Stillinger}},
  \bibinfo{journal}{Phys. Rev. B} \textbf{\bibinfo{volume}{32}},
  \bibinfo{pages}{5402} (\bibinfo{year}{1985}{\natexlab{b}}).

\bibitem[{\citenamefont{B{\"u}chner and Heuer}(1999)}]{buchner1999potential}
\bibinfo{author}{\bibfnamefont{S.}~\bibnamefont{B{\"u}chner}} \bibnamefont{and}
  \bibinfo{author}{\bibfnamefont{A.}~\bibnamefont{Heuer}},
  \bibinfo{journal}{Phys. Rev. E} \textbf{\bibinfo{volume}{60}},
  \bibinfo{pages}{6507} (\bibinfo{year}{1999}).

\bibitem[{\citenamefont{Broderix et~al.}(2000)\citenamefont{Broderix,
  Bhattacharya, Cavagna, Zippelius, and Giardina}}]{broderix2000energy}
\bibinfo{author}{\bibfnamefont{K.}~\bibnamefont{Broderix}},
  \bibinfo{author}{\bibfnamefont{K.~K.} \bibnamefont{Bhattacharya}},
  \bibinfo{author}{\bibfnamefont{A.}~\bibnamefont{Cavagna}},
  \bibinfo{author}{\bibfnamefont{A.}~\bibnamefont{Zippelius}},
  \bibnamefont{and} \bibinfo{author}{\bibfnamefont{I.}~\bibnamefont{Giardina}},
  \bibinfo{journal}{Phys. Rev. Letters} \textbf{\bibinfo{volume}{85}},
  \bibinfo{pages}{5360} (\bibinfo{year}{2000}).

\bibitem[{\citenamefont{Oligschleger and
  Schober}(1999)}]{oligschleger1999collective}
\bibinfo{author}{\bibfnamefont{C.}~\bibnamefont{Oligschleger}}
  \bibnamefont{and} \bibinfo{author}{\bibfnamefont{H.~R.}
  \bibnamefont{Schober}}, \bibinfo{journal}{Phys. Rev. B}
  \textbf{\bibinfo{volume}{59}}, \bibinfo{pages}{811} (\bibinfo{year}{1999}).

\bibitem[{\citenamefont{La~Nave et~al.}(2003)\citenamefont{La~Nave, Sciortino,
  Tartaglia, Shell, and Debenedetti}}]{la2003test}
\bibinfo{author}{\bibfnamefont{E.}~\bibnamefont{La~Nave}},
  \bibinfo{author}{\bibfnamefont{F.}~\bibnamefont{Sciortino}},
  \bibinfo{author}{\bibfnamefont{P.}~\bibnamefont{Tartaglia}},
  \bibinfo{author}{\bibfnamefont{M.~S.} \bibnamefont{Shell}}, \bibnamefont{and}
  \bibinfo{author}{\bibfnamefont{P.~G.} \bibnamefont{Debenedetti}},
  \bibinfo{journal}{Phys. Rev. E} \textbf{\bibinfo{volume}{68}},
  \bibinfo{pages}{32103} (\bibinfo{year}{2003}).

\bibitem[{\citenamefont{Likos}(2001)}]{likos2001eis}
\bibinfo{author}{\bibfnamefont{C.}~\bibnamefont{Likos}},
  \bibinfo{journal}{Phys. Rep.} \textbf{\bibinfo{volume}{348}},
  \bibinfo{pages}{267} (\bibinfo{year}{2001}).

\bibitem[{\citenamefont{Ashcroft and Mermin}(1976)}]{ashcroft1976ssp}
\bibinfo{author}{\bibfnamefont{N.~W.} \bibnamefont{Ashcroft}} \bibnamefont{and}
  \bibinfo{author}{\bibfnamefont{N.~D.} \bibnamefont{Mermin}},
  \emph{\bibinfo{title}{{Solid State Physics}}} (\bibinfo{publisher}{Saunders
  College: Philadelphia, Pa}, \bibinfo{year}{1976}).

\bibitem[{\citenamefont{Torquato et~al.}(2011)\citenamefont{Torquato, Zachary,
  and Stillinger}}]{torquato2011duality}
\bibinfo{author}{\bibfnamefont{S.}~\bibnamefont{Torquato}},
  \bibinfo{author}{\bibfnamefont{C.~E.} \bibnamefont{Zachary}},
  \bibnamefont{and} \bibinfo{author}{\bibfnamefont{F.~H.}
  \bibnamefont{Stillinger}}, \bibinfo{journal}{Soft Matter}
  \textbf{\bibinfo{volume}{7}}, \bibinfo{pages}{3780} (\bibinfo{year}{2011}).

\bibitem[{\citenamefont{Stillinger and Weber}(1982)}]{stillinger1982hidden}
\bibinfo{author}{\bibfnamefont{F.~H.} \bibnamefont{Stillinger}}
  \bibnamefont{and} \bibinfo{author}{\bibfnamefont{T.~A.} \bibnamefont{Weber}},
  \bibinfo{journal}{Phys. Rev. A} \textbf{\bibinfo{volume}{25}},
  \bibinfo{pages}{978} (\bibinfo{year}{1982}).

\bibitem[{\citenamefont{Shah and Chakravarty}(2002)}]{shah2002potential}
\bibinfo{author}{\bibfnamefont{P.}~\bibnamefont{Shah}} \bibnamefont{and}
  \bibinfo{author}{\bibfnamefont{C.}~\bibnamefont{Chakravarty}},
  \bibinfo{journal}{Phys. Rev. Lett.} \textbf{\bibinfo{volume}{88}},
  \bibinfo{pages}{255501} (\bibinfo{year}{2002}).

\bibitem[{\citenamefont{Qi et~al.}(2010)\citenamefont{Qi, Wang, Han, and
  Chen}}]{qi2010melting}
\bibinfo{author}{\bibfnamefont{W.~K.} \bibnamefont{Qi}},
  \bibinfo{author}{\bibfnamefont{Z.}~\bibnamefont{Wang}},
  \bibinfo{author}{\bibfnamefont{Y.}~\bibnamefont{Han}}, \bibnamefont{and}
  \bibinfo{author}{\bibfnamefont{Y.}~\bibnamefont{Chen}}, \bibinfo{journal}{J.
  Chem. Phys.} \textbf{\bibinfo{volume}{133}}, \bibinfo{pages}{234508}
  (\bibinfo{year}{2010}).

\bibitem[{\citenamefont{Mattarelli et~al.}(2010)\citenamefont{Mattarelli,
  Gasperi, Montagna, and Verrocchio}}]{mattarelli2010transparency}
\bibinfo{author}{\bibfnamefont{M.}~\bibnamefont{Mattarelli}},
  \bibinfo{author}{\bibfnamefont{G.}~\bibnamefont{Gasperi}},
  \bibinfo{author}{\bibfnamefont{M.}~\bibnamefont{Montagna}}, \bibnamefont{and}
  \bibinfo{author}{\bibfnamefont{P.}~\bibnamefont{Verrocchio}},
  \bibinfo{journal}{Phys. Rev. B} \textbf{\bibinfo{volume}{82}},
  \bibinfo{pages}{094204} (\bibinfo{year}{2010}).

\bibitem[{\citenamefont{Mattarelli et~al.}(2007)\citenamefont{Mattarelli,
  Montagna, and Verrocchio}}]{mattarelli2007ugc}
\bibinfo{author}{\bibfnamefont{M.}~\bibnamefont{Mattarelli}},
  \bibinfo{author}{\bibfnamefont{M.}~\bibnamefont{Montagna}}, \bibnamefont{and}
  \bibinfo{author}{\bibfnamefont{P.}~\bibnamefont{Verrocchio}},
  \bibinfo{journal}{Appl. Phys. Lett.} \textbf{\bibinfo{volume}{91}},
  \bibinfo{pages}{061911} (\bibinfo{year}{2007}).

\bibitem[{\citenamefont{Torquato et~al.}(2006)\citenamefont{Torquato, Uche, and
  Stillinger}}]{torquato2006rsa}
\bibinfo{author}{\bibfnamefont{S.}~\bibnamefont{Torquato}},
  \bibinfo{author}{\bibfnamefont{O.~U.} \bibnamefont{Uche}}, \bibnamefont{and}
  \bibinfo{author}{\bibfnamefont{F.~H.} \bibnamefont{Stillinger}},
  \bibinfo{journal}{Phys. Rev. E} \textbf{\bibinfo{volume}{74}},
  \bibinfo{pages}{61308} (\bibinfo{year}{2006}).

\bibitem[{\citenamefont{S{\"u}t{\H{o}}}(2005)}]{suto2005cgs}
\bibinfo{author}{\bibfnamefont{A.}~\bibnamefont{S{\"u}t{\H{o}}}},
  \bibinfo{journal}{Phys. Rev. Lett.} \textbf{\bibinfo{volume}{95}},
  \bibinfo{pages}{265501} (\bibinfo{year}{2005}).

\bibitem[{\citenamefont{Verlet}(1967)}]{verlet1967cec}
\bibinfo{author}{\bibfnamefont{L.}~\bibnamefont{Verlet}},
  \bibinfo{journal}{Phys. Rev.} \textbf{\bibinfo{volume}{159}},
  \bibinfo{pages}{98} (\bibinfo{year}{1967}).

\bibitem[{\citenamefont{Dennis and Mei}(1979)}]{dennis1979tnu}
\bibinfo{author}{\bibfnamefont{J.~E.} \bibnamefont{Dennis}} \bibnamefont{and}
  \bibinfo{author}{\bibfnamefont{H.~H.~W.} \bibnamefont{Mei}},
  \bibinfo{journal}{J. Optim. Theory Appl.} \textbf{\bibinfo{volume}{28}},
  \bibinfo{pages}{453} (\bibinfo{year}{1979}).

\bibitem[{\citenamefont{Kaufman}(1999)}]{kaufman1999rsq}
\bibinfo{author}{\bibfnamefont{L.}~\bibnamefont{Kaufman}},
  \bibinfo{journal}{SIAM J. Optim.} \textbf{\bibinfo{volume}{10}},
  \bibinfo{pages}{56} (\bibinfo{year}{1999}).

\bibitem[{\citenamefont{Henkelman and
  J{\'o}nsson}(2000)}]{henkelman2000improved}
\bibinfo{author}{\bibfnamefont{G.}~\bibnamefont{Henkelman}} \bibnamefont{and}
  \bibinfo{author}{\bibfnamefont{H.}~\bibnamefont{J{\'o}nsson}},
  \bibinfo{journal}{J. Chem. Phys.} \textbf{\bibinfo{volume}{113}},
  \bibinfo{pages}{9978} (\bibinfo{year}{2000}).

\bibitem[{\citenamefont{Henkelman et~al.}(2002)\citenamefont{Henkelman,
  J{\'o}hannesson, and J{\'o}nsson}}]{henkelman2002methods}
\bibinfo{author}{\bibfnamefont{G.}~\bibnamefont{Henkelman}},
  \bibinfo{author}{\bibfnamefont{G.}~\bibnamefont{J{\'o}hannesson}},
  \bibnamefont{and}
  \bibinfo{author}{\bibfnamefont{H.}~\bibnamefont{J{\'o}nsson}},
  \bibinfo{journal}{Theoret. Methods  Condensed Phase Chem.} pp.
  \bibinfo{pages}{269--302} (\bibinfo{year}{2002}).

\bibitem[{\citenamefont{Stillinger}(1999)}]{stillinger1999exponential}
\bibinfo{author}{\bibfnamefont{F.~H.} \bibnamefont{Stillinger}},
  \bibinfo{journal}{Phys. Rev. E} \textbf{\bibinfo{volume}{59}},
  \bibinfo{pages}{48} (\bibinfo{year}{1999}).

\end{thebibliography}
\end{document}